\documentclass{pasj01}
\Received{$\langle$21-May-2024$\rangle$}
\Accepted{$\langle$24-Aug-2024$\rangle$}
\Published{$\langle$publication date$\rangle$}
\newcommand{\degr}{^\circ}

\usepackage[switch,mathlines]{lineno}

\begin{document}

\title{Survey of Non-thermal Electron around Supermassive Black Holes through Polarization Flips}
\author{Yuh Tsunetoe,\altaffilmark{1,2}$^*$
        Tomohisa Kawashima,\altaffilmark{3}
        Ken Ohsuga,\altaffilmark{2}
        Shin Mineshige\altaffilmark{4}}%
\altaffiltext{1}{Black Hole Initiative, Harvard University, \ 20 Garden street, Cambridge, \ MA 02138, USA}
\altaffiltext{2}{Center for Computational Sciences, University of Tsukuba, \ 1-1-1 Tennodai, Tsukuba, \ Ibaraki 305-8577, Japan}
\altaffiltext{3}{Institute for Cosmic Ray Research, University of Tokyo, \ 5-1-5 Kashiwanoha, Kashiwa, \ Chiba 277-8582, Japan}
\altaffiltext{3}{Department of Astronomy, Kyoto University, Kitashirakawa-Oiwake-cho, Sakyo-ku, \ Kyoto 606-8502, Japan}
\email{ytsunetoe@fas.harvard.edu}

\KeyWords{polarization -- radiative transfer -- radiation mechanisms: non-thermal -- galaxies: jets -- black hole physics}

\maketitle

\begin{abstract}
Optically thick non-thermal synchrotron sources notably produce linear polarization vectors being {\it parallel} to projected magnetic field lines on the observer's screen, although they are {\it perpendicular} in well-known optically thin cases.
To elucidate the complex relationship between the vectors and fields and to investigate the energy and spatial distribution of non-thermal electrons through the images, we perform polarization radiative transfer calculations at submillimeter wavelengths. 
Here the calculations are based on semi-analytic force-free jet models with non-thermal electrons with a power-law energy distribution. 
In calculated images, we find a $90\degr$-flip of linear polarization (LP) vectors at the base of counter-side (receding) jet near a black hole, which occurs because of large optical depths for synchrotron self-absorption effect. 
The $90\degr$-flip of LP vectors is also seen on the photon ring at a high frequency, since the optical depth along the rays is large there due to the light bending effect. 
In addition, we see the flip of the sign of circular polarization (CP) components on the counter jet and photon ring. 
Furthermore, we show that these polarization flips are synthesized with large values in the spectral index map, and also give rise to outstanding features in the Faraday Rotation Measure (RM) map. 
Since the conditions of flipping depend on the magnetic field strength and configuration and the energy distribution of electrons, we can expect that the polarization flips will provide us with an observational evidence for the presence of non-thermal electrons around the black hole, and a clue to the magnetically driving mechanism of plasma jets. 

\end{abstract}

%\pagewiselinenumbers

\section{Introduction}

Synchrotron radiations transport fruitful informations concerning magnetic field configuration and other emitting plasma properties in forms of  polarization components, spectral energy distributions, and so on (e.g., \cite{1959ApJ...130..241W,1965ARA&A...3..297G}). 
In particular, linear polarization (LP) vectors from synchrotron sources have attracted a great deal of interest and been vigorously observed for a diversity of astronomical targets, due to their characteristics imprinting the directions of magnetic fields. 

For example, it is well-known in optically thin cases that LP vectors are aligned perpendicular to the directions of magnetic field lines projected on the transverse plane along light rays, if they are described in the electric vector position angle (EVPA) notation (e.g., \cite{1979rpa..book.....R}). 
In these cases, circular polarization (CP) components imprints the polarity of magnetic fields (e.g., \cite{1968ApJ...154..499L}). 

Meanwhile, it has been pointed out in literatures that the relationship between polarization components and magnetic fields can be changed under the existence of synchrotron self-absorption (SSA) effect (e.g., \cite{1967ApJ...150..647P}); that is, in optically thick cases with non-thermal electron distributions, LP vectors in the EVPA are converged into those parallel with the direction of projection of magnetic field lines on observer's screen, which are different by $90\degr$ from those in the optically thin cases (\cite{1970ApJ...161...19A}; see also \cite{2018Galax...6....5W} for a recent review). 
Simultaneously, CP components also change their signs, from left (right) to right (left) handedness \citep{1971Ap&SS..12..172M,1977ApJ...214..522J}. 

These counterintuitive behavior is triggered by imbalance between the polarized synchrotron emission and polarized SSA effect.
In such cases, the synchrotron electrons work like a polarizing filter shutting out the components aligned with the direction of their motions. 
As a result, they produce the flipping polarization components originated from the unpolarized component of synchrotron emission (see Subsection \ref{subsec:eqs} for detail). 

These features of synchrotron plasmas should be noted in that one must interpret each pixel of LP vector maps in two different ways, depending on their optically thickness for the SSA. 
Actually, in observations of the active galactic nucleus (AGN) jets, the existence of significant SSA effect has been suggested from the core shifts (e.g., \cite{2011Natur.477..185H,2012A&A...545A.113P}) and large spectral indices at radio frequencies (e.g., \cite{2014AJ....147..143H}). 

Conversely, we can also think of utilizing the flipping polarization components to survey the composition of particles in the AGNs and their jets. 
\citet{1970ApJ...161...19A} interpreted time-variable LP vector angles of quasars as results of variations in the optical thickness of the sources. 
\citet{2000ApJ...538L.121A} adopted a uniform slab model with two components of  power-law and thermal electrons, to reproduce an observed $90\degr$-flip in LP position angle of Sgr A* in submillimeter spectra \citep{2000ApJ...534L.173A}. 

In this work, we investigate the feasibility of verifying the non-thermal particle distribution through the polarization flipping features on images. 
Recent observations by very long baseline interferometers (VLBI) have been accessible to the innermost, base regions of AGN jets close to the central supermassive black holes (SMBH) (e.g., \cite{2019ApJ...875L...1E,2023Natur.616..686L}), in which the essential accretion and creation processes of plasma particles are thought to take place. 
To probe the relationship between the non-thermal particles and polarimetric features on the images, we perform general relativistic radiative transfer (GRRT) calculations based on semi-analytic jet models with a power-law distribution of particles. 

\bigskip

The organization of this paper is outlined as follows: 
Section \ref{subsec:model} outlines the methodology to obtain theoretical polarization images using jet fluid models and GRRT. 
Section \ref{subsec:eqs} presents a review of the flipping polarization components in the optically thick non-thermal plasmas, using a simplified radiative transfer equation. 
Moving forward, Section \ref{sec:result} focuses on a resultant image, to elucidate how the linear and circular polarization flips can occur and be observed on the innermost jet base and on the photon ring. 
Here, we also compare the polarization features with the spectral index map. 
Furthermore, Section \ref{sec:discussion} addresses the prospects and limitations of our results, for example, effects of thermal accretion disk, relationship with the rotation measure (RM) maps, and application to a diversity of AGN jets. 

\section{Jet Model and Radiative Transfer}\label{subsec:model}

\subsection{Force-Free Jet}

We assume the force-free solution to the magnetohydrodynamic (MHD) equations with the magnetically dominated regime in mind. 
Both of numerical simulations and observational estimations have suggested that such situations can be applied to funnel jet regions above a SMBH and accretion disk (e.g., \cite{2009MNRAS.394L.126M,2015ApJ...803...30K}).

The GR-covariant form of the force-free plasma jet models has been suggested to reproduce the ring-like images around the black hole at submillimeter wavelengths \citep{2009ApJ...697.1164B}, and the limb-brightened and triple-ridged features of M87 jet at millimeters \citep{2018ApJ...868...82T,2019ApJ...877...19O}. 
\citet{2018ApJ...868...82T} applied the force-free jet models to the description based on the Blandford-Znajek process \citep{1977MNRAS.179..433B}, by threading the magnetic field lines into the event horizon. 

\bigskip

The force-free jet models are outlined as follows. 
The magnetic field is given by a stream function with a parameter $\nu$ in the spherical coordinates $(r, \theta, \phi)$, 
\begin{equation}
	\Phi = B_{0}~r^{\nu_{\rm st}}(1\mp{\rm cos}~\theta) \ \ (- \ {\rm for} \ z\ge0, \ + \ {\rm for} \ z<0), 
	\label{eq:stream}
\end{equation}
which forms conical and parabolic jet for $\nu_{\rm st} = 0$ and $\nu_{\rm st} = 1$, respectively. 
Then the poloidal and toroidal components of magnetic field are described as follows, if in a 3-dimensional form, 
\begin{eqnarray}
	{\bf B}_{p} = \frac{\nabla \Phi \times \hat{\phi}}{R}, \\
	B_{\phi} = \mp \frac{2\Omega_{F}\Phi}{Rc},
\end{eqnarray}
respectively, where $c$ is the speed of light, $(R,\phi,z)$ are the cylindrical coordinates, and $\Omega_{F}$ is the angular velocity of the field. 

The plasma velocity is assumed to be the drift velocity, 
\begin{equation}
	{\bf v} = \frac{{\bf E} \times {\bf B}}{B^{2}}c, 
\end{equation}
in which the electric field is given as ${\bf E} = \Omega_{F}\hat{\phi} \times {\bf B}$. 
Under such a condition, the magnetic and velocity field configurations can be demarcated by the light cylinder radius $R_{\rm lc}$, defined by $R_{\rm lc}\Omega_{F}/c = 1$. For $R \ll R_{\rm lc}$ and $R \gg R_{\rm lc}$, the magnetic fields are dominated by the poloidal and toroidal  components, respectively. 
Meanwhile, the fluid velocity becomes toroidal-dominant and sub-relativistic (poloidal-dominant and relativistic) for $R \ll R_{\rm lc}$ ($R \gg R_{\rm lc}$) \citep{2018ApJ...868...82T}.

Furthermore, the assumption of the continuity equation of plasma density gives a constant $n/B^{2}$ along the field line. 
Then, by choosing a Gaussian distribution $n = n_{0}~{\rm exp}~(-r^{2}{\rm sin}^{2}~\theta/ 2r_{\rm fp}^{2})$ at a footprint $|r{\rm cos}~\theta| = r_{\rm fp}$, the particle number density is given in each point along the field line, with a supplementary factor $\{1-{\rm exp}~(-r^{2}/2r_{\rm fp}^{2})\}$ mocking the mass loading above a black hole. 
The expressions in the covariant form and more detail properties can be found in the references above.

\bigskip

We set $\nu_{\rm st}=1$ in Eq.~\ref{eq:stream} and consider the magnetic fields threaded into the event horizon as in \citet{2018ApJ...868...82T}, putting 
\begin{equation}
	\Omega_{F} = \frac{ac}{4r_{+}},
\end{equation}
where $a$ is the spin parameter of black hole and $r_{+} = (1+\sqrt{1-a^{2}})r_{g}$ is the radius of event horizon. Here $r_{g} = GM/c^{2}$ is the gravitational radius, $G$ is the gravitational constant and $M$ is the mass of black hole. 
We also take $a = 0.5~M$ and $M = 6.5\times10^{9}M_{\odot}$, where $M_{\odot}$ is the Solar mass. 
In this case, the light cylinder radius is located at $R_{\rm lc} = 14~r_{\rm g}$.

\begin{figure*}
\begin{center}
	\includegraphics[width=15cm]{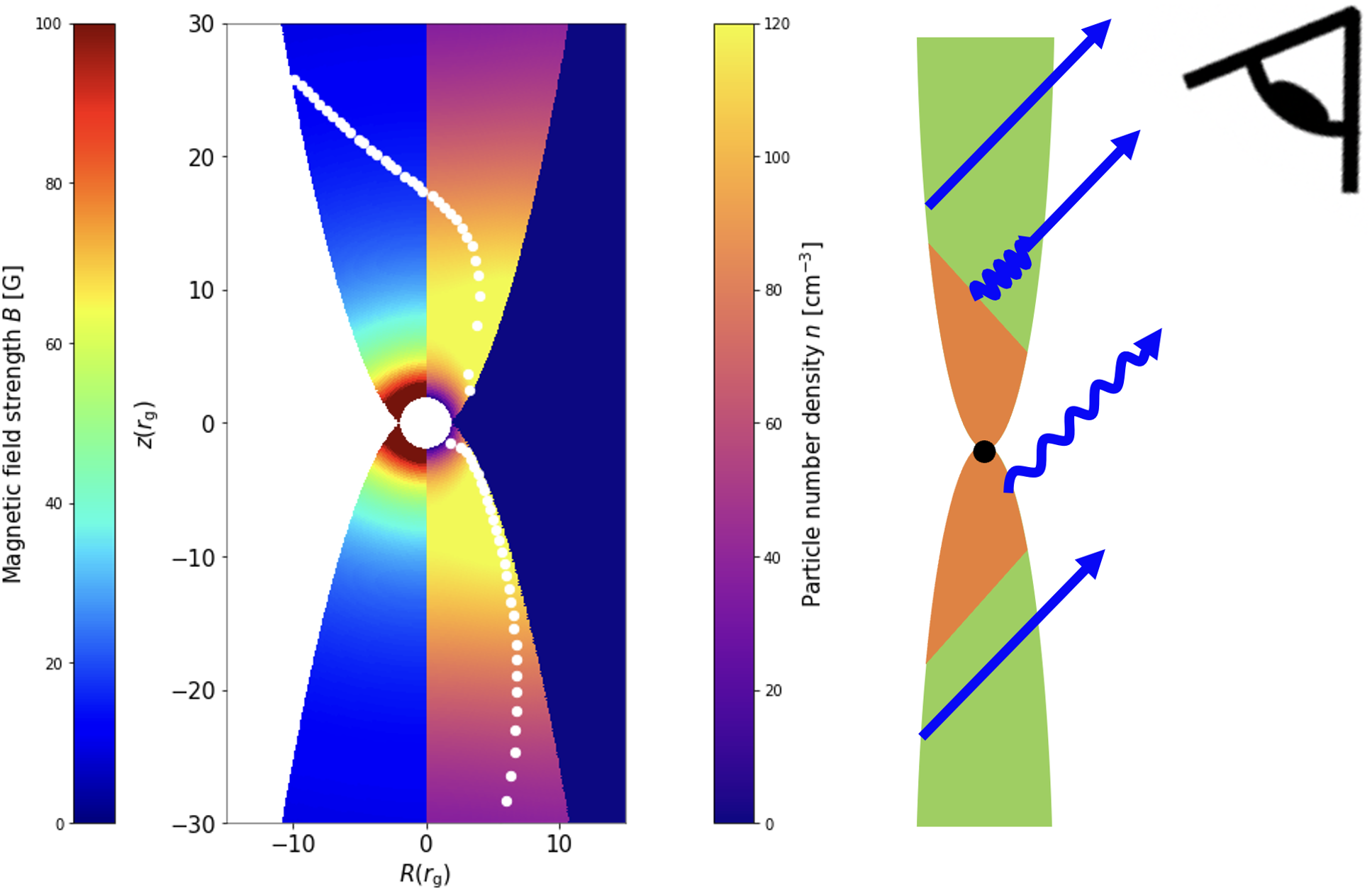}
\end{center}
    \caption{Left: Poloidal slice maps at $\phi = 0$ of the nonthermal particle density, magnetic field strength.
    The white points are the SSA photosphere at 43~GHz of $\tau_{\rm SSA} = 1$ for light paths of the pixels on the $x=0$ vertical line in the right image of Fig.~\ref{fig:43Gjet}. 
    The observer's screen is located in the top-right direction with inclination angle of $i = 20\degr$ to the $z$-axis. 
    All the points are projected onto the poloidal slice of $\phi=0$. 
    Right: Picture of radiative transfer from the jets. The intrinsic (straight arrow) and flipped (wavy arrow) polarization components are radiated from the optically thin jets (green) and the photosphere (orange), respectively. The flipped components from the counter jet reach to the observer, while those from the approaching jet can be drowned by the optically-thin emission components in the outer,  downstream regions. 
    The light bending effect is ignored in the picture.
    }
    \label{fig:map}
\end{figure*}

The magnetic fields are scaled with $B_{0}$ in Eq.~\ref{eq:stream}, taken as $300~{\rm G}$ with an innermost region of AGN jets in mind here. 
The nonthermal particle density is scaled by $n_{0}$ with $r_{\rm fp}= 10r_{\rm g}$, so that it gives total flux value of $\sim 1-2~{\rm Jy}$ at millimeter to submillimeter wavelengths at the distance of M87. 
In addition, we put $n=0$ for the magnetic fields not threaded into the horizon, that is, $\Phi > B_{0}~r_{+}$, as in \citet{2019ApJ...877...19O}. 
The poloidal slice maps of the magnetic field strength $B$ and particle density $n$ are shown in Fig.~\ref{fig:map}. 

Finally, we assume that the nonthermal electrons take an isotropic, power-law distribution described by Eq.~\ref{eq:power-law}. 
The power-law index, minimum and maximum Lorentz factors are set to $p=3.5$, $\gamma_{\rm min} = 30$ and $\gamma_{\rm max} = 10^{6}$, respectively.

\subsection{General Relativistic Polarization Radiative Transfer}

We implement the force-free jet model into our polarimetric GRRT code \texttt{SHAKO}\footnote{Its name refers to \textit{Oratosquilla oratoria}, a kind of mantis shrimps, in Japanese. They perceive both linearly and circularly polarized lights, and create shockwaves with their punches.} (\texttt{S}tokes-based \texttt{H}orizon-scale \texttt{A}strophysical \texttt{K}nowledge \texttt{O}utputter), which was firstly implemented in \citet{2020PASJ...72...32T}. 
In Appendix \ref{app:codecomp}, we perform a code comparison with an existing code, \texttt{ipole} \citep{2018MNRAS.475...43M,2022ApJS..259...64W}, and confirm good agreement in calculated polarization images based on general relativistic magnetohydrodynamics (GRMHD) fluid model.

The polarized emissivities of synchrotron emission and radiative coefficients of SSA for the power-law distribution are implemented into the code by tabulating integrals of modified Bessel functions \citep{1959ApJ...130..241W}, rather than by adopting the approximate forms with Gamma functions \citep{1968ApJ...154..499L}, which should be invalid at frequencies in a range of $\nu \lesssim \nu_{\rm c}$, of our interest (see \citet{2016MNRAS.462..115D} for comparison between these two; see also Appendix \ref{app:spectra} in this paper for deviations from the approximate expressions in Eq.~\ref{eq:thickLPf}). 

The coefficients of Faraday rotation and conversion are referred to from the approximate expressions in \citet{1969SvA....13..396S} and \citet{1977ApJ...214..522J}, which work well for our $\gamma_{\rm min} = 30$ \citep{2011MNRAS.416.2574H}. 
We survey a competition among the SSA and Faraday effects in Appendix \ref{app:spectra}.

\bigskip

We set the observer's screen at a distance of $r = 10^{4} r_{\rm g}$ with an inclination angle of $i = 20\degr$ to the $z$-axis. 
All the model parameters are fixed to the values above throughout this paper except for Subsection \ref{subsec:diverse}, in which we discuss a more powerful jet case with larger magnetic field strength and plasma density. 
Although our parameter setting is based on the estimated values in M87, we aim, in the present study, to discuss a comprehensive prospects for polarization images with a diversity of AGN jets in mind.

\section{Analytical Expression of Polarization Flip by SSA}\label{subsec:eqs} 

While we solve the full set of polarimetric GRRT equations in this paper, we here demonstrate the effects of the flip of LP components by ignoring the contribution of CP component, Faraday rotation of LP vector, and the relativistic effects. 

In this simplified way, the polarimetric radiative transfer equations for synchrotron plasma are written as follows:
\begin{equation}
	\frac{\rm d}{{\rm d}s}
	\left(
		\begin{array}{c}
			I \\
			Q \\
		\end{array}
	\right)
		=
	\left(
		\begin{array}{c}
			j_I \\
			j_Q \\
		\end{array}
	\right)
		-
	\left(
		\begin{array}{cc}
			\alpha_{I} & \alpha_{Q} \\
			\alpha_{Q} & \alpha_{I} \\
		\end{array}
	\right)
	\left(
		\begin{array}{c}
			I \\
			Q \\
		\end{array}
	\right), 
	\label{eq:radtransIQ}
\end{equation}
where $I,Q$ are Stokes parameters for the specific total intensity and LP component, respectively, and $j_{I,Q}$ and $\alpha_{I,Q}$ are the synchrotron emissivities and SSA coefficients. 
Now we omit Stokes $U$, another parameter for LP component, by taking the bases of polarization reference (sky) plane so that $\alpha_{U}=0$,\footnote{This corresponds to taking one of the bases parallel with projected magnetic field component on the screen.} which also follows $j_{U}=0$ for synchrotron radiation. 

Under the assumption of constant coefficients, the LP fraction in the optically thick limit, $\tau_{\rm SSA} \equiv \int\alpha_I {\rm d}s \gg 1$, is converged to  
\begin{equation}
    \left(\frac{Q}{I}\right)_{\rm thick} = \frac{j_Q \alpha_I - j_I \alpha_Q}{j_I \alpha_I - j_Q \alpha_Q},   
	\label{eq:QoI}
\end{equation}
for arbitrary electron-energy distributions, which we deduce for a more general case with Faraday rotation in Appendix \ref{app:analysis}. 

In the case of thermal electrons, the emissivities and absorption coefficients follow the Kirchoff's law for each polarized component, $j_{(I,Q)} = \alpha_{(I,Q)}B_{\nu}$, respectively, where $B_{\nu}$ is the Planck function (see, for example, \cite{1979rpa..book.....R}). 
Substituting into Eq.~\ref{eq:QoI}, these give $(Q/I)_{\rm thick} = 0$; that is, synchrotron radiations from an optically thick thermal plasma are perfectly unpolarized.  

\bigskip

We here consider a non-thermal, power-law energy distribution, 
\begin{equation}
    N\left(\gamma\right) = 
    	\left\{
		\begin{array}{cc}
			n \frac {p-1}{\gamma_{\rm min}^{1-p} - \gamma_{\rm max}^{1-p}} \gamma^{-p} \ &\left(\gamma_{\rm min} \le \gamma \le \gamma_{\rm max}\right) \\
			0 \ &\left({\rm otherwise}\right)
		\end{array}
    	\right.,
	\label{eq:power-law}
\end{equation}
where $n$ is the particle number density, $\gamma$ is the Lorentz factor of particles, and $p$, $\gamma_{\rm min}$ and $\gamma_{\rm max}$ are the power-law index, minimum and maximum Lorentz factors, respectively. 
At a range of $\nu_{\rm c} \ll \nu \ll \nu_{\rm c}(\gamma_{\rm max}/\gamma_{\rm min})^{2}$, this yields 
\begin{equation}
	j_{Q} = \frac{p+1}{p+(7/3)} j_{I}, \\
	\label{eq:jLPf}
\end{equation}
\begin{equation}
	\alpha_{Q} = \frac{p+2}{p+(10/3)} \alpha_{I},
	\label{eq:alphaLPf}
\end{equation}
where $\nu_{\rm c} \equiv (3/2)\nu_{B}{\rm sin}\,\theta_{B}\gamma_{\rm min}^{2}$, $\nu_{B} \equiv eB/m_{\rm e}c$ and $\theta_{B}$ is the angle between light ray and magnetic field line (see, for example, \cite{1980gbs..bookQ....M}). 
Here $e$ is the charge of electron, $B$ is strength of magnetic field, $m_{\rm e}$ is the mass of electron, and $c$ is the speed of light. 
In the optically thick limit, one obtain 
\begin{equation}
    \left(\frac{Q}{I}\right)_{\rm thick} = - \frac{3}{6p+13}, 
    \label{eq:thickLPf}
\end{equation}
which has an opposite sign to that in the well-known optically thin limit, 
\begin{equation}
	\left(\frac{Q}{I}\right)_{\rm thin} = \frac{j_{Q}}{j_{I}} = \frac{p+1}{p+(7/3)}. 
\end{equation}
Since the sign of $Q$ determines whether the LP vector is vertical or horizontal on the polarization plane, this means that LP vectors from an optically thick plasma should flip by $90\degr$ in their position angles, with respect to those from an optically thin plasma. 

\bigskip

\begin{figure*}
\begin{center}
	\includegraphics[width=15cm]{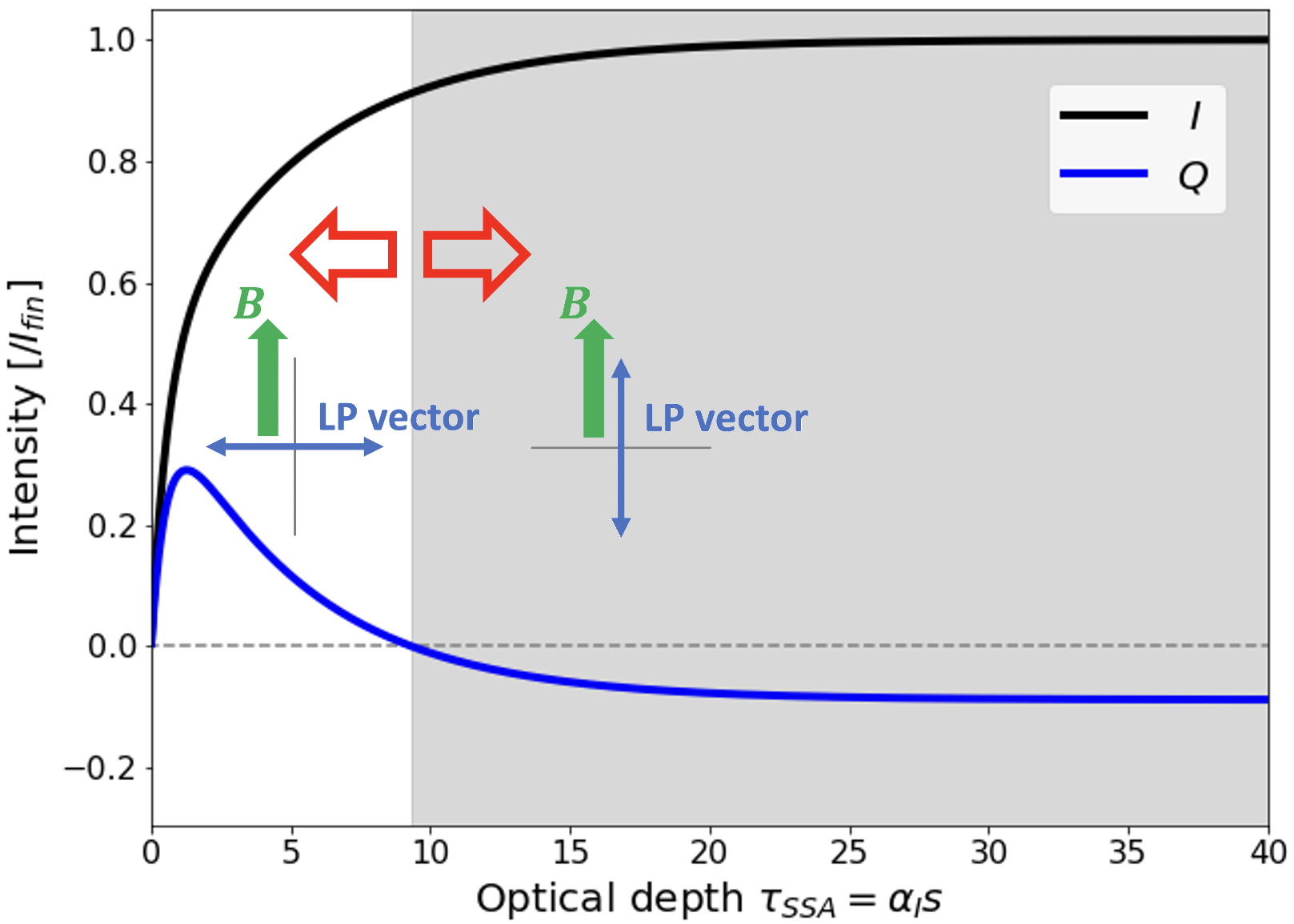}
\end{center}
    \caption{A diagram displaying the propagation of Stokes $I$ and $Q$ in Eq.~\ref{eq:radtransIQ} produced by non-thermal particles under constant coefficients satisfying Eqs.~\ref{eq:jLPf} \& \ref{eq:alphaLPf}.
    We take the optical depth for the SSA effect $\tau_{\rm SSA}$ as the horizontal axis. The vertical axis is the Stokes $I$ or $Q$ normalized with the final, converged value of the total intensity, $I_{\rm fin}$. 
    Stokes parameter $I$ is the total intensity and always positive, while $Q$ is the LP component and can be negative. 
    Here the power-law index is set to $p=3.5$. 
    Two inset pictures show the relationship between the LP vector and projected magnetic field line ${\it B}$ on the polarization plane, in the EVPA notation. The optically thick region in which the LP vector is flipped by $90\degr$ is colored with gray. 
    }
    \label{fig:IQrt}
\end{figure*}

To see the behavior of LP vector from non-thermal synchrotron plasma introduced above, we plot radiative transfer propagation of Eq.~\ref{eq:radtransIQ} under the conditions of Eqs.~\ref{eq:jLPf} \& \ref{eq:alphaLPf} with $p = 3.5$ in Figure \ref{fig:IQrt}. 

As expected, $I$ and $Q$ increase linearly in the early stage of $\tau_{\rm SSA} \ll 1$ with a LP fraction $\sim +80\%$, which corresponds to a LP vector perpendicular to the projected magnetic field on the polarization plane in EVPA. 

Propagating in the plasma, the light become subject to significant SSA effect, $\tau_{\rm SSA} \sim 1$. 
Then the LP component $Q$ begins to stagnate and decay relatively to the total intensity $I$. 
The decreasing $Q$ changes its sign at $\tau_{\rm SSA} \sim 10$, which means that the LP vector angle is flipped\footnote{Hereafter, we adopt a terminology of {\it flip}, rather than ``rotation'', to avoid a confusion with the Faraday rotation.} by $90\degr$ and become parallel with the projection of magnetic field line in EVPA. 

Finally, $I$ and $Q$ converge into a LP fraction $\sim -10\%$ in an optically thick limit $\tau_{\rm SSA} \gtrsim 30$.
The convergence of the total and LP intensities is delayed by the existence of absorption of polarization components, compared to the unpolarized case (see Appendix \ref{app:analysis}). 

\bigskip

We can also apply the above discussion to the CP component by changing $Q$ into $V$ in Eqs.~\ref{eq:radtransIQ} \& \ref{eq:QoI}. 
In this case, the sign-change in Stokes $V$ corresponds to the transition of handedness of CP, which imprints the polarity of magnetic field in the optically thin case. 
The matter is a little complicated in that the emissivity and absorption coefficient of CP are not written in the simple way as Eqs.~\ref{eq:jLPf} \& \ref{eq:alphaLPf} (see \cite{1971Ap&SS..12..172M} for tabulated CP fractions in the optically thick limit). 
Typically, the flipping CP component converges into $(V/I)_{\rm thick} \sim 0.1\%$, an order of magnitude lower than the CP fraction in the optically thin case. 

\bigskip

So far we showed that the LP vector (and CP component) flips in the optically thick, non-thermal plasma with SSA effect. 
Meanwhile, in our calculation, all the LP and CP components are transferred in parallel being subject to Faraday rotation and conversion, in addition to the SSA. 
Furthermore, millimeter and submillimeter wavelengths, which is of our interest, can be out of the range $\nu_{\rm c} \ll \nu \ll \nu_{\rm c}(\gamma_{\rm max}/\gamma_{\rm min})^{2}$, the assumption for Eqs.~\ref{eq:jLPf} \& \ref{eq:alphaLPf}.

As will be shown in the results in Section \ref{sec:result} and later, the flipping LP vectors on the images are distributed on the images with a wide range of fractions, $\sim 0 - 10\%$, depending the optical depths and the factors mentioned in the last paragraph.
In Appendices \ref{app:analysis} and \ref{app:spectra}, we discuss the applicability of flipping polarizations at multi-wavelengths by surveying a competition among the SSA and two Faraday effects, and a deviation from Eqs.~\ref{eq:jLPf}, \ref{eq:alphaLPf}, and thus Eq.~\ref{eq:thickLPf}.

\section{Resultant Images}\label{sec:result}

%\subsection{Toy Model with Constant Coefficients}
%
%To check the applicability of the description in subsection \ref{subsec:eqs} to realistic situations, we here introduce a toy model with constant physical values to perform radiative transfer. 
%With typical values estimated for the innermost region of M87 jet in mind, we assume the magnetic field strength $B = 30~{\rm G}$, the number density of non-thermal electrons $n_{\rm nth} = 1~{\rm cm^{-3}}$, and the power-law index $p = 3.5$ and cutoff Lorentz factors, $(\gamma_{\rm min},\gamma_{\rm max}) = (30, 10^6)$. For simplicity, we consider a plasma at rest, ignoring the GR effects. 
%
%In radiative transfer calculation, we keep $\theta_B = 45\degr$, the angle between magnetic field and line of sight, constant. 

%\subsection{Jet Polarization Images at 43~GHz}

\subsection{$90\degr$-Flip of LP Vectors on Counter-side Jet}\label{subsec:jetLPreversal}

\begin{figure*}
\begin{center}
	\includegraphics[width=17.5cm]{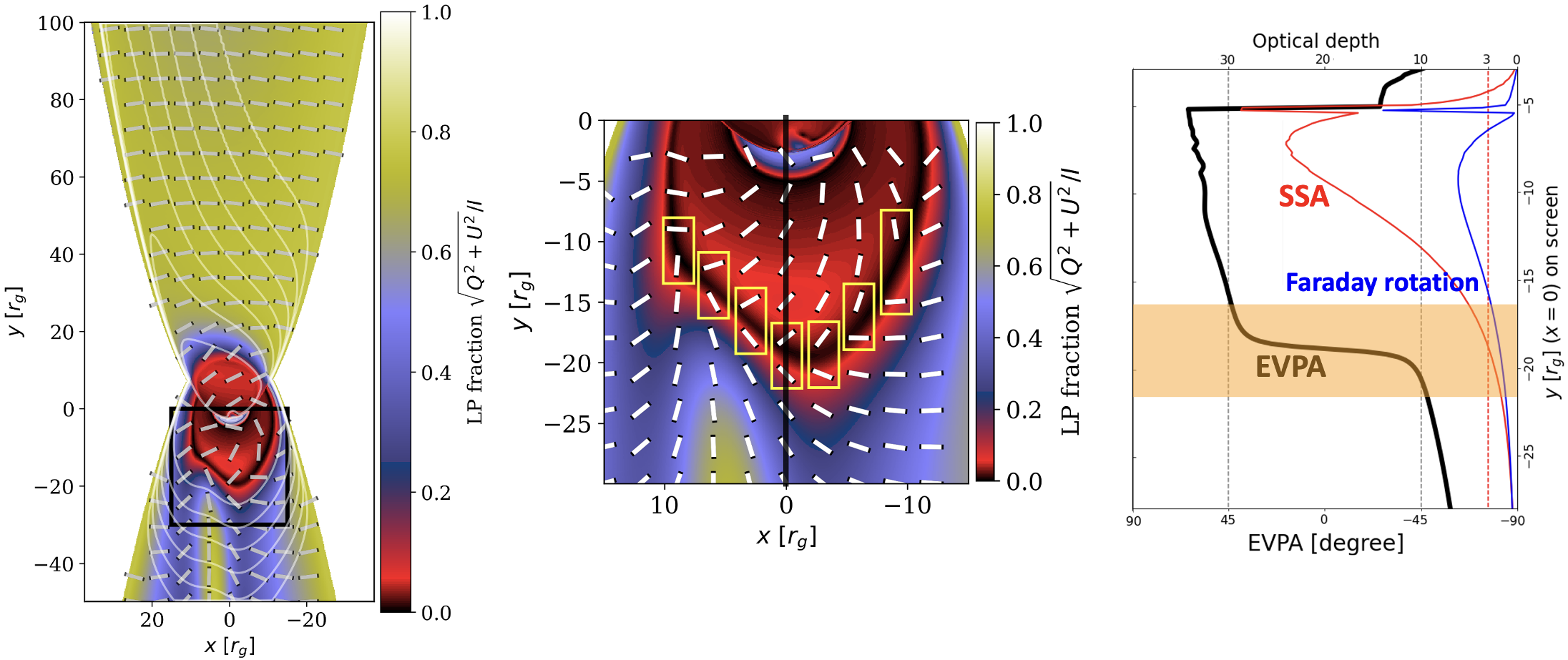}
\end{center}
    \caption{Left: LP map at 43~GHz and its zooming up into the boxed region (center). 
    The line contour, color contour, and ticks correspond to the total intensity, LP fraction, and LP vectors in EVPA, respectively. 
    The $z$-axis in the coordinates (and thus the angular momentum vector of black hole) points upwards in projection onto the images. 
    Center: Zooming-up image into the boxed region in the left image. Pairs of the flipping LP vectors are shown in the yellow boxes. 
    Right: Vertical profile of the EVPA (black solid line and bottom axis) , and optical depths for SSA and Faraday rotation for each light path (red and blue solid lines and top axis) along $(0,y)$ in the screen coordinates (black line in the central panel). The vector angle is measured with respect to $x$-axis of the screen and is positive for counter-clockwise direction. Two gray, dotted lines corresponds to ${\rm EVPA} = \pm45\degr$. A red dotted line points $\tau = 3$. 
    The range of the yellow box in the central panel is colored with orange. 
    }
    \label{fig:43Gjet}
\end{figure*}

The calculated LP maps at 43~GHz are shown in two panels in Fig.~\ref{fig:43Gjet}, in which the total intensity, LP fraction, and LP vectors in EVPA are shown by line contour, color contour, and ticks, respectively. 

In the left panel, the foreground-side, approaching (or counter-side, receding) jet extends upwards (downwards). 
While the foreground jet gives an globally ordered vector pattern as a whole, the counter jet shows dis-ordered vectors in both of vertical and horizontal direction (even without SSA or Faraday effects; see Appendix \ref{app:noSSA}). 
Except the flip features introduced below, these tendencies are roughly consistent with those in \citet{2009ApJ...697.1164B}, in spite of the quite different choice of $\Omega_{F}$ and a moderate spin value of black hole.\footnote{They survey two spin values of $a=0.998$ and $0$ with their M0 and M2 models, respectively. Our vector pattern is relatively close to the latter.} 

\bigskip

In the base region of the counter jet at $r \lesssim 20~r_{\rm g}$, as is indicated by a box in the left panel of Fig.~\ref{fig:43Gjet}, we can find an irregular feature; the LP fractions drop and rise from the downstream to upstream of jet, crossing the line of zero LP. In addition, the LP vector angles drastically change in the vertical direction across which the zero LP points. 
If we zoom up this region (see the central panel), we can clearly see that the angle of LP vectors changes by $90\degr$ in the vertical direction with a ditch-like border of zero-LP fraction, as shown by each pair of ticks in the yellow boxes. 

Furthermore, it is confirmed in the right panel that the LP vector position angles along the central, vertical line of $(0,y)$ on the screen are actually flipped from $\sim -45\degr$ to $\sim +45\degr$ in a border of $y \approx -18~r_{\rm g}$ with the SSA optical depth of $\tau_{\rm SSA} \approx 3$. The flipping is brought forward compared to that in Fig.~\ref{fig:IQrt} due to subdominant, but significant Faraday rotation (see Appendix \ref{app:analysis}). 
Here, we note that the optical depth is computed along each light path, so that it is neither monotonically increasing or decreasing function on $y$.

There is another feature of the flip of LP vector around $y \approx -5 r_{\rm g}$ on the right panel, until where the EVPA profile smoothly changes. 
Here, the optical depth profiles have spikes corresponding to the photon ring (which cannot be seen clearly in the total and LP intensities), on which light rays gain larger path lengths by the light bending (see also subsection \ref{subsec:86Gring}). 
Inside the photon ring, the EVPA drops back to $\sim -45\degr$, or re-flips, with smaller optical depths for light rays plunging into the event horizon. 

\bigskip

As seen above, the flipping LP vectors appear in the counter jet base, in which the SSA effect becomes significant due to large nonthermal electron density and magnetic field strength. 

In order to confirm the relationship between the flipping and optical depth, we show the location of the photosphere ($\tau_{\rm SSA} = 1$) of the global jets at the observed frequency $\nu = 43~{\rm GHz}$ with the points in Fig.~\ref{fig:map}. 
Here, each point is calculated for light path for each pixel with $x=0$ (so composing the central vertical line) in the left image of Fig.~\ref{fig:43Gjet}, and is projected onto the poloidal slice of $\phi = 0$.\footnote{Note that here the optical depth is calculated reversely to the radiative transfer, from the screen to the object.} 
Here $\tau_{\rm SSA} \equiv \int \alpha_{I}{\rm d}\lambda$ is the optical depth for SSA effect, and $\lambda$ is the affine parameter in light path propagation.

The photosphere is located along the region close to the jet border (or sheath) at the base of counter jet, while it is located inside of the jet in the approaching and downstream of counter jet. 
The former shows the polarization components flipped inside the photosphere. 
However, in the latter case, the components that are once flipped are drowned by the intrinsic, non-flipped emissions in the foreground, optically-thin jet region. 
As a result, the base of approaching jet shows low LP fractions but no flip of vectors (In subsection \ref{subsec:diverse}, the flipping in the approaching side is discussed). 

Fig.~\ref{fig:map} also shows the SSA photosphere in the counter jet base is located in a region with magnetic field strengths $B \sim 10 - 100~{\rm G}$, which give the synchrotron characteristic frequencies $\nu_{\rm c} \sim 10- 100~{\rm GHz}$ for $\gamma_{\rm min} = 30$. 
This result is consistent with the fact that the polarization flip can occur at frequencies $\nu \sim \nu_{\rm c} $, at which the SSA effect peaks out and can dominate over Faraday rotation and conversion, as shown in Appendices \ref{app:analysis} and \ref{app:spectra}. 
Conversely, we can expect to survey the strength and configuration of magnetic fields, in addition to the non-thermal particle distribution, in the jet base through the polarization flips. 

\bigskip

We surveyed the force-free jet models for a conical streamline case $\nu_{\rm st} = 0$ and for low and high black hole spins, $a = 0.1, 0.998$, with all the other parameter values being unchanged. As a result, we confirmed the occurrence of the polarization flipping in the counter-side jet in all the models.

Furthermore, it is confirmed that the flipping region is more extended downwards at lower frequency, down to $y \sim -100 r_{\rm g}$ ($\sim -0.4$ mas for M87) at 4 GHz. We also find the flipping in the approaching-side jet at 4, 8 and 15 GHz, as will also be seen in the high-inclination case in subsection \ref{subsec:diverse}.

\subsection{Flipped LP vectors on Photon Ring}\label{subsec:86Gring}

\begin{figure*}
\begin{center}
	\includegraphics[width=17.5cm]{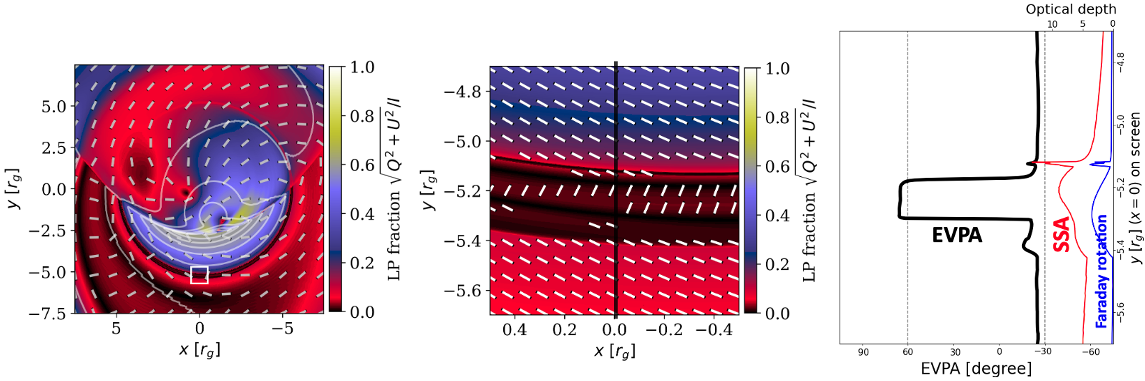}
\end{center}
    \caption{LP map at 86~GHz (left), its zooming up onto the boxed region on the photon ring (center), and vertical profiles of EVPA and optical depths for SSA and Faraday rotation along $(0,y)$ on the image, the black line in the central panel (right).
    }
    \label{fig:86Gring}
\end{figure*}

As mentioned in the last section, we can also expect to see the polarization flipping features on the photon ring, on which the consisting light rays gain large optical depths due to the light bending effect. In order to show more prominent feature near the photon ring in $\sim 0.1~r_{\rm g}$ scale, we introduce the LP map at a higher frequency,  86~GHz at which the optical depth is smaller on the overall image. 

The LP map at 86~GHz and its zooming-up into a part of photon ring are shown in the left and central panels in Fig.~\ref{fig:86Gring}, respectively.
Since the plasma becomes optically thinner for the SSA at higher frequencies, the photon ring feature can be clearly seen in the left panel. 
The ring gives low LP fractions because of the combination of large optical depths for SSA and the Faraday effects, rotation of polarization plane along bent light paths (gravitational Faraday rotation), and turbulent magnetic fields (see \cite{2021MNRAS.503.4563J}; see also Subsection \ref{subsec:ringLPvec}). 

However, a part of the ring shows the relatively strong LP components reversed by $90\degr$ to the inside and outside of the ring as shown in the right panel, forming a dim, $n=1$ sub-ring feature on the dark, ditch-like photon ring.\footnote{A narrower $n=2$ subring can also be seen above the $n=1$ ring.} (Here, the $n$-th subring on the photon ring consists of light rays on $n$ half orbits around the black hole.) 
This consists of light rays gaining large SSA optical depths dominant over those for Faraday rotation and conversion, satisfying the condition of polarization flipping as shown in Appendix \ref{app:spectra}. 
These can be a clue to the non-thermal particles and magnetic fields in the innermost plasmas close to the black hole.

\subsection{Flip of CP components}

\begin{figure*}
\begin{center}
	\includegraphics[width=15cm]{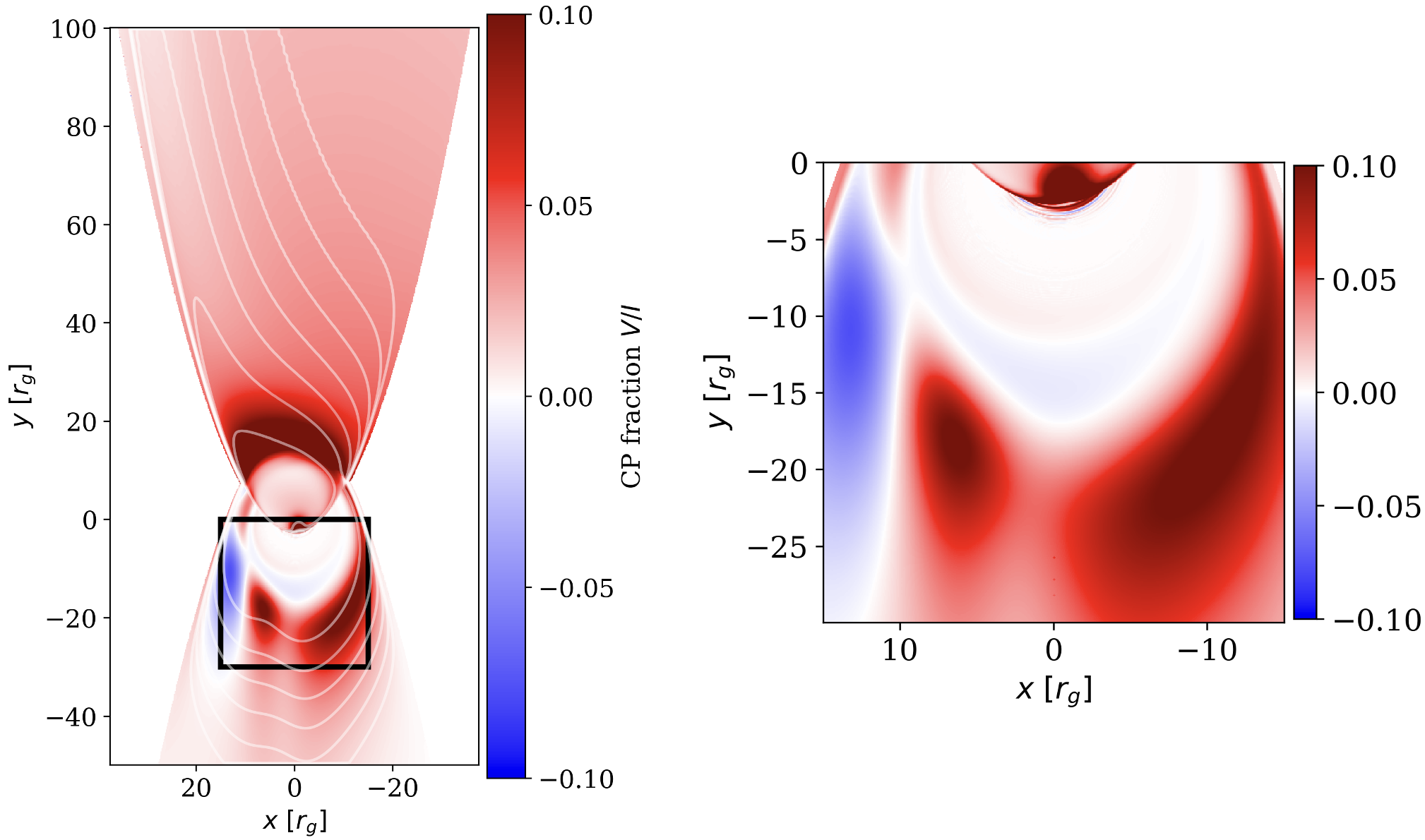}
\end{center}
    \caption{CP image at 43~GHz and its zooming up. The color and line contours denote the CP and total intensities, respectively. 
    The square box in Fig.~\ref{fig:43Gjet} is overlaid for comparison.
    }
    \label{fig:43GjetCP}
\end{figure*}

\begin{figure*}
\begin{center}
	\includegraphics[width=15cm]{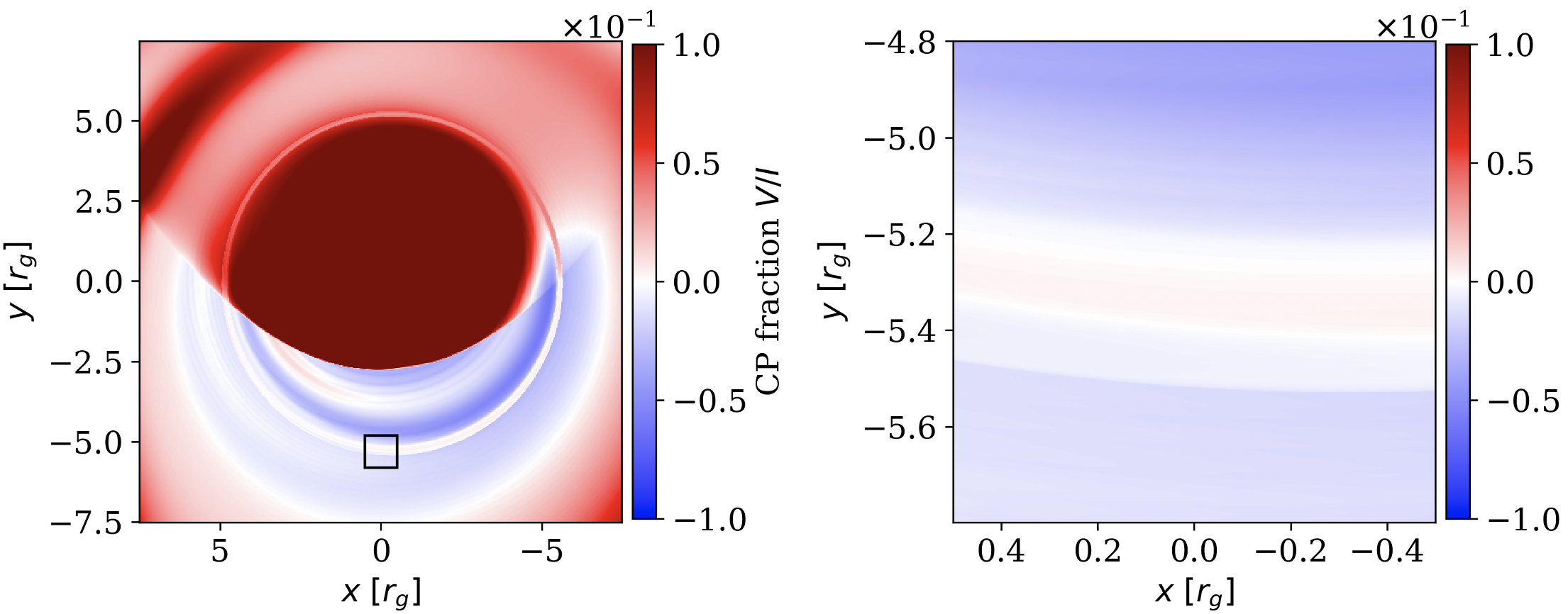}
\end{center}
    \caption{CP map at 86~GHz and its zooming up.
    }
    \label{fig:86GringCP}
\end{figure*}

The CP images at 43 and 86~GHz are shown in Figs.~\ref{fig:43GjetCP} \& \ref{fig:86GringCP}, which are the counterparts to the LP maps in Figs.~\ref{fig:43Gjet} \& \ref{fig:86Gring}, respectively. 
In the base regions of jets, the CP components can be increased through Faraday conversion process in relativistic plasmas up to $\sim 10\%$ (see \cite{2020PASJ...72...32T,2021MNRAS.505..523R}). 
Thus, we here examine the flip of CP components with the effect of Faraday conversion in mind. 

In Fig.~\ref{fig:43GjetCP}, we can see dim ($\sim 0.5\%$ in absolute fraction), negative CP components at $y \sim -15~r_{\rm g}$ in the base of counter jet, although the flipping region is narrower than that in the LP map in Fig.~\ref{fig:43Gjet} and is dominated over and drowned by the positive components in the innermost region of $y \lesssim -10~r_{\rm g}$. 
The drowning positive CP components are explained by a dominant Faraday conversion effect. 
As shown in Fig.~\ref{fig:alphas_spectrum} in Appendix, Faraday conversion becomes stronger than SSA and Faraday rotation at low frequencies, $\nu \ll \nu_{\rm c}$. 
Thus, the CP components at 43~GHz become weak but non-reversed in the innermost region with magnetic field strengths of $B \gtrsim 50~{\rm G}$, which corresponds to $\nu_{\rm c} \gtrsim 100~{\rm GHz}$. 

\bigskip

The flipped CP components are also seen on the photon ring in the left panel of Fig.~\ref{fig:86GringCP}, as the sub-ring of weak positive ($\sim 0.3\%$) components. 
It should be noted here that the CP components by the intrinsic emissions are negative in the lower (counter jet's) side of the innermost region.

\bigskip

As a whole, the flip of CP components is more fragile to the Faraday effects than that of LP vectors, since their weaker components at the emissions can be easily drowned by Faraday-converted  components. 
In addition, we should note that the reversed components on the photon ring can also be given arise to by another effect (see Subsection \ref{subsec:ringLPvec}). 
With this caution in mind, we still expect that the flipping CP components will be used as a supplementary evidence for nonthermal plasmas in combination with the LP vector flips.

\subsection{Comparison with Spectral Index Map}

\begin{figure*}
\begin{center}
	\includegraphics[width=15cm]{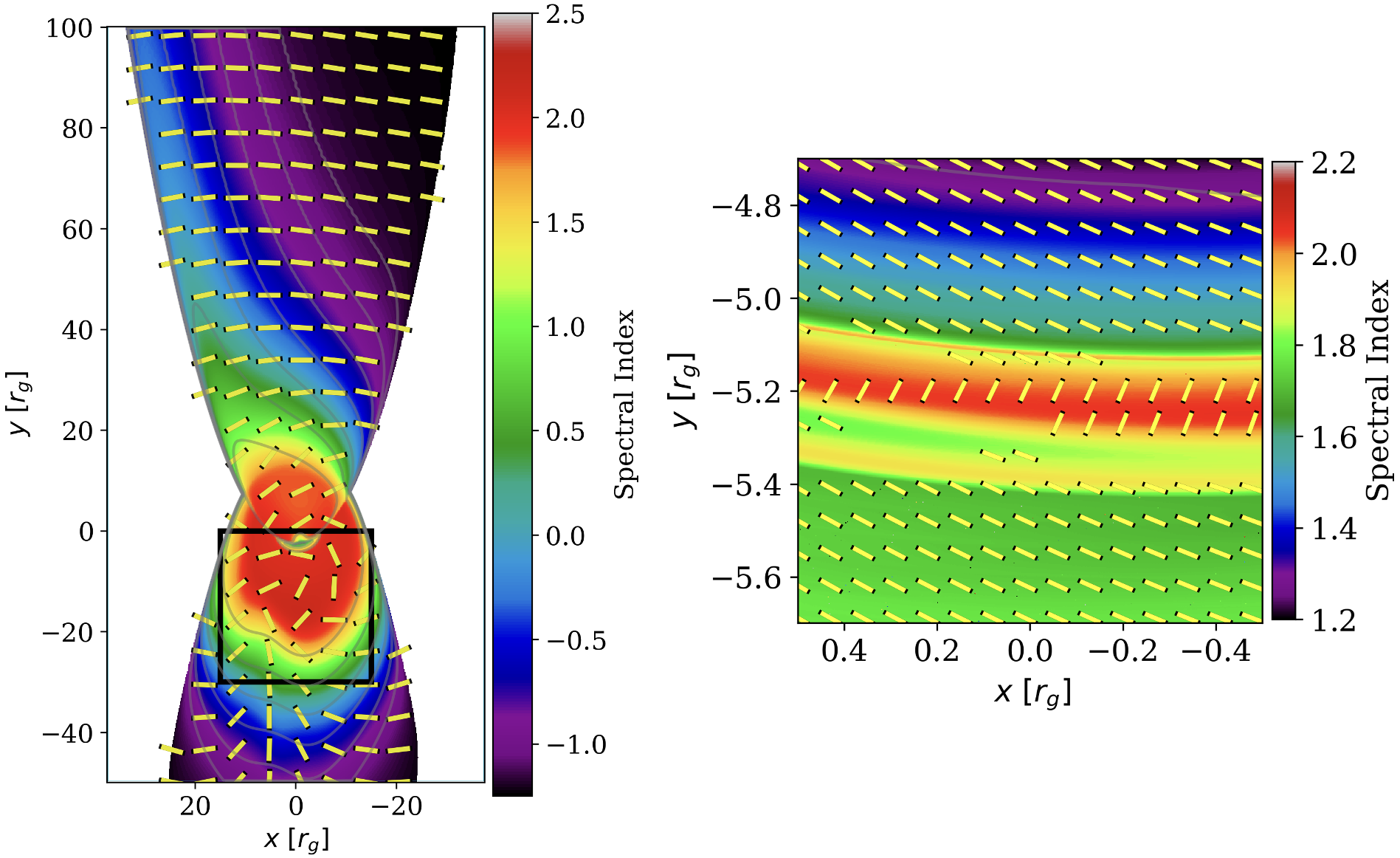}
\end{center}
    \caption{Spectral index maps of the jet and on the photon ring, calculated from two intensities in each pixel at 22-43~GHz (left) and 84-86~GHz (right). 
    The ticks of LP vectors and line contour of total intensity are referred to from the left panel of Fig.~\ref{fig:43Gjet} (left) and the right panels of Fig.~\ref{fig:86Gring} (right). 
    In the left panel, the square in Fig.~\ref{fig:43Gjet} is overplotted for comparison. 
    Note that the scales of color contour are different in two maps. 
    }
    \label{fig:43Gjet_spindex}
\end{figure*}

In relation to the flipping polarizations by SSA effect, let us compare with the map of spectral index $\alpha$, where we assume the total specific intensity at a frequency $\nu$ to be $I_{\nu} \propto \nu^{\alpha}$. 
It should be noted in the optically thick limit that the total intensity of radiation from polarized sources generally converges into a value different from that of the source function $S_{\nu} = j_{I}/\alpha_{I}$. 
For example, Eq.~\ref{eq:radtransIQ} with constant coefficients has the analytic solution converging into, 
\begin{equation}
	I \rightarrow \frac{j_{I}\alpha_{I}-j_{Q}\alpha_{Q}}{\alpha_{I}^{2}-\alpha_{Q}^{2}}, 
\end{equation}
which is not equal to $j_{I}/\alpha_{I}$ unless $\alpha_{Q} = 0$. 
Substituting Eqs.~\ref{eq:jLPf} and \ref{eq:alphaLPf} into this, in fact, we obtain $I \approx 1.076j_{I}/\alpha_{I}$ for $p=3.5$, which still holds $I \propto j_{I}/\alpha_{I}$. 

Thus, the well-known values of $\alpha = -(p-1)/2$ in the optically thin limit and $\alpha = 2.5$ in the optically thick limit for the power-law distribution (e.g., \cite{1979rpa..book.....R}) are also applied to polarized sources in the range of $\nu_{\rm c} \ll \nu \ll \nu_{\rm c}(\gamma_{\rm max}/\gamma_{\rm min})^{2}$ (see also Appendix \ref{app:analysis} for the expressions including Faraday rotation).

Two spectral index maps of the jet and on the photon ring (zooming-up) are shown in the left and right panels in Fig.~\ref{fig:43Gjet_spindex}, which are calculated from two total intensities at 22-43~GHz and 84-86~GHz, respectively. 
As expected, the spectral indices are negative, down to $-(p-1)/2 = -1.25$, in the optically thin downstreams of twin jets in the left panel, while they increase and become positive going to the optically thick upstreams. 
In the approaching jet, the left side of the jet shows larger $\alpha$, since the SSA effect is also relativistically beamed by rotational motion of plasmas, as the emission.

In both panels, the large values $\alpha > 2$ are synchronized with the flips of LP vectors. 
Such large values are characteristic to the power-law electrons, compared to $\alpha = 2$ in the optically thick limit for the thermal electrons. 
Therefore, we can expect to verify the non-thermal electrons close to the black hole through survey of these two features on the unpolarized and polarized images.

\section{Discussion}\label{sec:discussion}

\subsection{Thermal Particles in Accretion Disk}
\label{subsec:thermaldisk}

\begin{figure*}
\begin{center}
	\includegraphics[width=10cm]{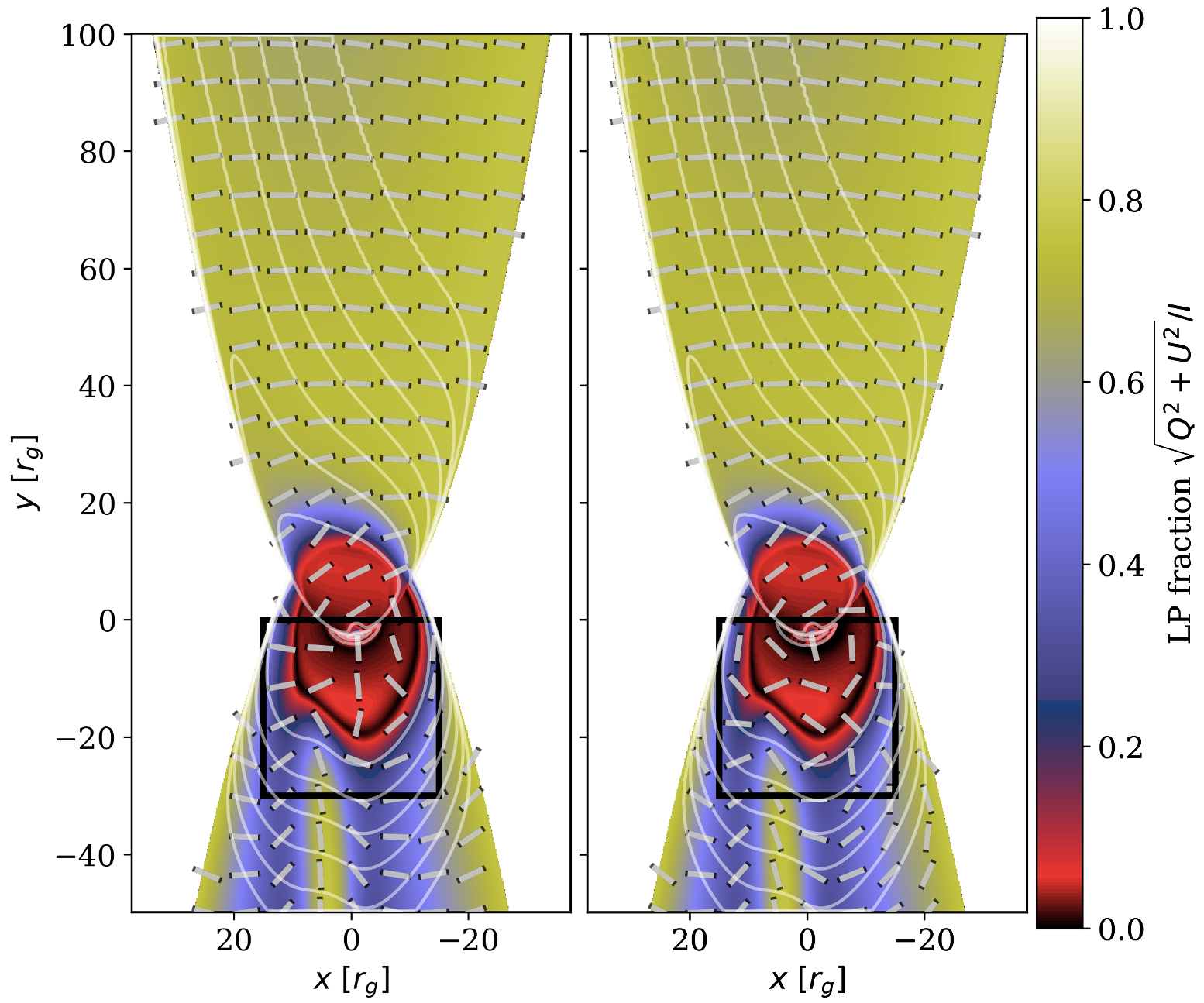}
\end{center}
    \caption{Same as the left panel of Fig.~\ref{fig:43Gjet}, but for the case, in which a thermal accretion disk is added.
    The disks are either moderately ($\beta=3$) or strongly ($\beta=0.3$) magnetized in the left and right panels, respectively. 
    The square box in Fig.~\ref{fig:43Gjet} is overplotted for comparison.}
    \label{fig:43Gjet_ring}
\end{figure*}

So far we assumed purely non-thermal plasmas as radiation sources, but there should be a thermal disk component in reality. 
To consider the effect of accretion disk on the polarization flips, we introduce a thermal disk model in addition to the non-thermal force-free jet. 
Following analytical, Keplerian shell models of the radiatively inefficient accretion flows (RIAF) introduced in previous works (e.g., \cite{2006ApJ...636L.109B,2016ApJ...831....4P,2019ApJ...878...27K}), we set an axi-symmetric disk, which is assumed to be Keplerian-rotating and infalling outside and inside of the innermost stable circular orbit (ISCO), respectively, with the profiles of  
\begin{eqnarray}
	n_{\rm th} &=& 1\times10^{5}\left(\frac{r}{r_{\rm g}}\right)^{-1.5}{\rm exp}~\left(-\frac{z^{2}}{2\cdot0.5^{2}}\right)~{\rm cm^{-3}}, \\
	T_{\rm e} &=& 7\times10^{10}\left(\frac{r}{r_{\rm g}}\right)^{-1}~{\rm K}, 
\end{eqnarray}
where $n_{\rm th}$ and $T_{\rm e}$ are the number density and temperature of thermal electrons. 
These profiles are based on the estimations in M87 (see, for example, \cite{2019ApJ...875L...5E}). 
The disk has a scale height of $H \sim 0.5~R$ where $R=r~{\rm cos}~\theta$. 
Magnetic fields is assumed to be purely toroidal in the disk\footnote{We confirmed that the results in this subsection are qualitatively unchanged for two disk models with purely poloidal magnetic fields, one with vertical and another with radial ones.} with the energy density profile in equipartition with the protons,  
\begin{equation}
	\frac{B^{2}}{8\pi} = \beta^{-1}n_{\rm th}m_{\rm p}c^{2}\left(\frac{6r}{r_{\rm g}}\right)^{-1}, 
\end{equation}
where $m_{\rm p}$ is the mass of proton and $\beta$ is a fraction parameter of the equipartition.

Fig.~\ref{fig:43Gjet_ring} shows the LP maps at 43~GHz incorporating the effect of thermal disk for the cases with the equipartition parameter $\beta =$ $3$ (moderate magnetization) and $0.3$ (strong magnetization) in the left and right panels, respectively. 
First of all, we can see the thermal disks does not so much affect on the image morphologies at the low frequency (i.e., the line and color contours on two images). 
In this sense, the two images are dominated by the emissions from jets. 

While the effect of thermal disk is subdominant in the total intensity image, it can give rise to difference in the LP vector angles according to the magnetic field strength. 
The LP vectors on the counter jet are significantly scrambled by Faraday rotation in the disk in both images, while we can still see the ditch-like feature in the LP fractions. 
In the left panel, the $90\degr$-flips of the two vectors bordering the ditch of color contour are preserved as a whole. 
In the right panel, however, the flipping becomes relatively indistinguishable due to larger Faraday rotations with stronger magnetic fields. 
In the next subsection, we explore additional features detectable even in such cases.

\subsection{Effects on Rotation Measure Map}

\begin{figure*}
\begin{center}
	\includegraphics[width=7.5cm]{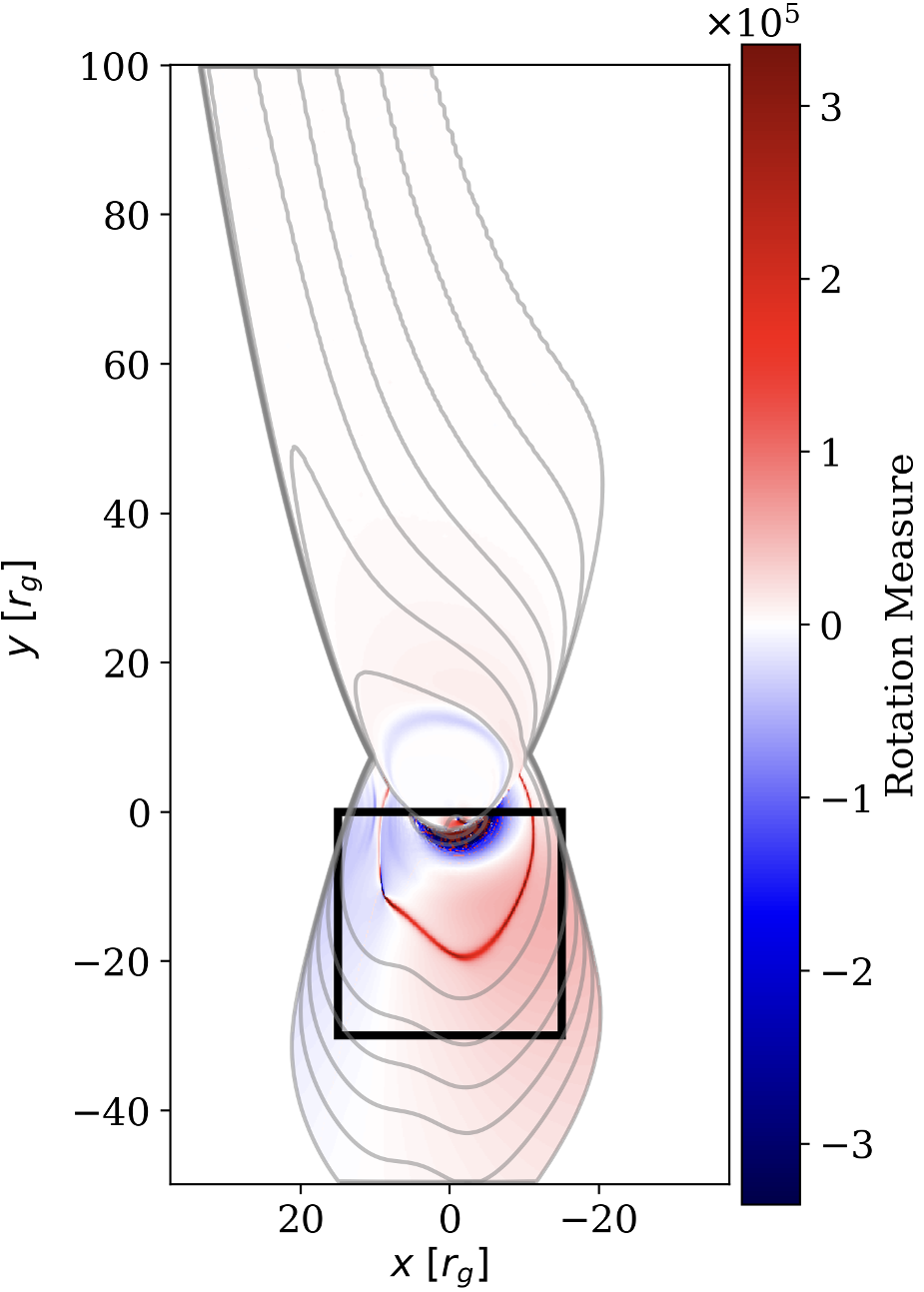}
\end{center}
    \caption{Rotation measure (RM) map calculated from two images at 42 and 43~GHz. The line contour expresses the total intensity at 43~GHz.
    The square box in Fig.~\ref{fig:43Gjet} is plotted for comparison.
    }
    \label{fig:43Gjet_RM}
\end{figure*}

One can expect that the  $90\degr$-flips of LP vectors should appear on the Faraday rotation measure (RM) maps, as well, but in some irregular way. 
RM is defined by 
\begin{equation}
	{\rm RM} \equiv \frac{\phi_{\nu_{1}}-\phi_{\nu_{2}}}{\lambda_{1}^{2}-\lambda_{2}^{2}}, 
\end{equation}
where $\phi_{\nu} = {\rm arg}(Q_{\nu}+iU_{\nu})/2$ is LP vector position angle and $\lambda = c/\nu$ is observational wavelength. 
In Fig.~\ref{fig:43Gjet_RM}, we show the RM map calculated from two images at 42 and 43~GHz for the model with strongly magnetized disk ($\beta=0.3$), which was introduced in the last subsection. 

On the map, large RM values rises up in the ditch region bordering the flipped and non-flipped LP vectors in the right panel of Fig.~\ref{fig:43Gjet_ring}. However, these are due to the SSA effect, not to the Faraday rotation. 
That is, one observes LP vectors rotated by $\approx 90\degr$ in the area on which the LP vectors are flipped by $90\degr$ at 42~GHz but not at 43~GHz. 
This is because the SSA effect becomes stronger so that the polarization flipping area extends outwards at lower frequencies. 

The large RM values on the flipping border are given as ${\rm RM} = (\pi/2+\Delta\phi_{\rm rot})/(\lambda_{1}^{2}-\lambda_{2}^{2})$, where $\Delta\phi_{\rm rot}$ is the angle offset by Faraday rotation, if the Faraday rotation occurs after the flipping by SSA. 
This could give rise to an ambiguity of $90\degr$ in the estimate of RM in the SSA thick region at multiwavelengths. 
This ambiguity can be resolved by comparison with the LP fraction, since the LP flipping and large RM by SSA is accompanied with a decrease of the LP fraction ($\lesssim 10\%$).

\subsection{Application to Diverse AGN jets}
\label{subsec:diverse}

\begin{figure*}
\begin{center}
	\includegraphics[width=16cm]{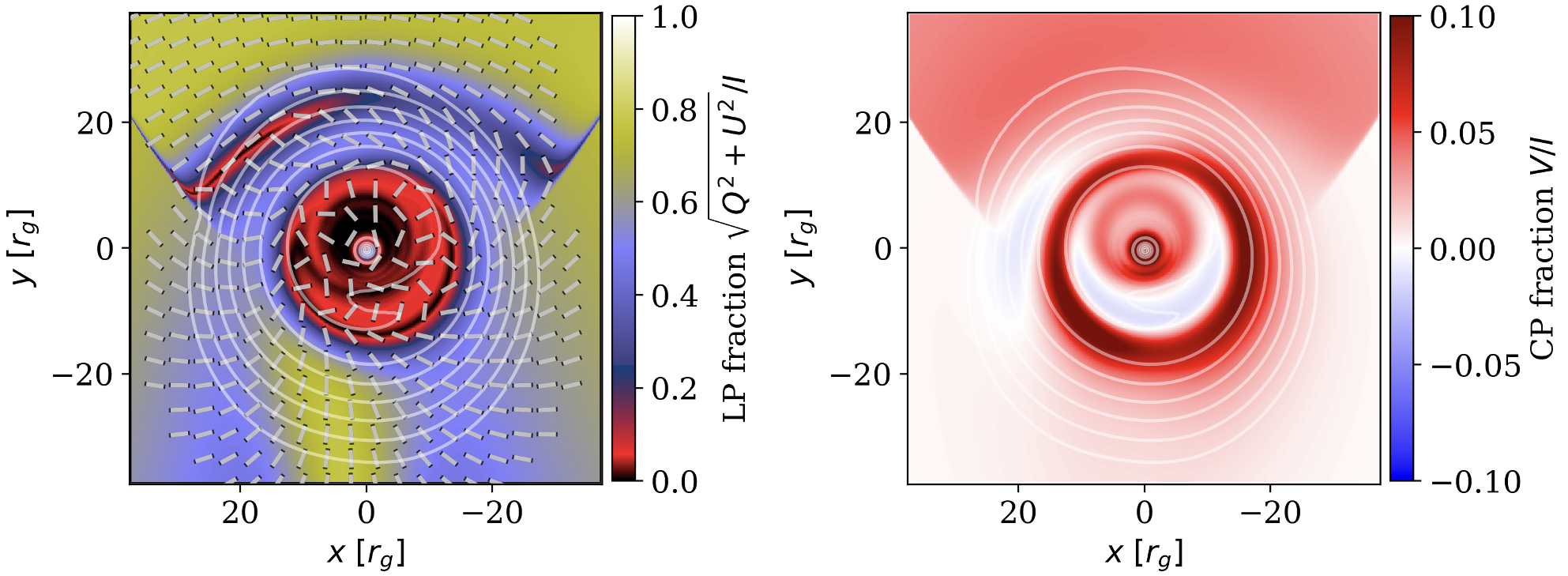}
\end{center}
    \caption{LP map and CP image at 230~GHz for a high density model with an inclination angle of $i = 5\degr$. }
    \label{fig:i5_LPCP}
\end{figure*}

\begin{figure*}
\begin{center}
	\includegraphics[width=16cm]{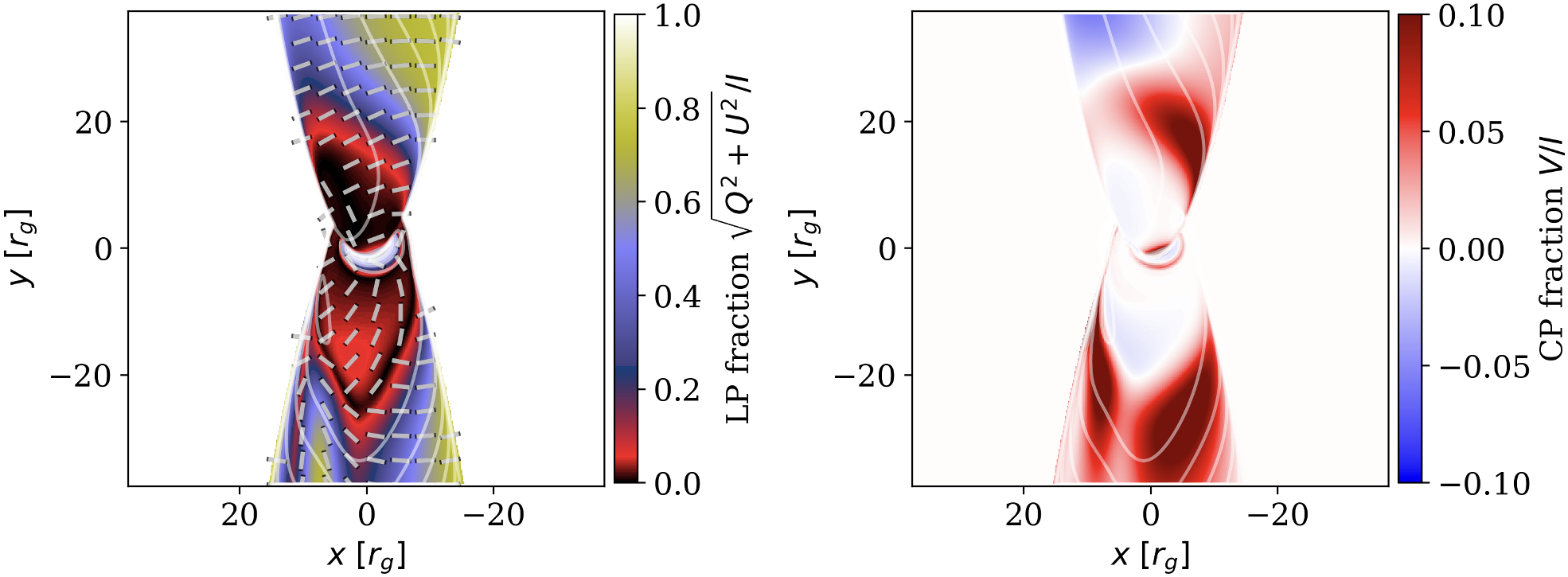}
\end{center}
    \caption{Same as Fig~\ref{fig:i5_LPCP}, but with $i = 45\degr$. }
    \label{fig:i45_LPCP}
\end{figure*}

Until now, we have discussed the polarization flipping for a force-free jet with the model parameters based on estimates in M87, which is categorized into the low-luminosity AGNs (LLAGNs; \cite{1997ApJS..112..315H}). 
Here, we extend the discussion to a diversity of AGN jets  with higher mass accretion rates.%, including quasars and high luminosity blazars, while the mass accretion rate expected by the examined parameter in this section is still lower than them. 

We increase the electron density and magnetic field strength by factors of  $10$ and $\sqrt{10}$, respectively, so that the mass accretion rate becomes 10 times higher with the plasma magnetization being unchanged. 
The LP and CP images at 230~GHz for this high accretion model with inclination angle of $5\degr$ and $45\degr$ are shown in Figs.~\ref{fig:i5_LPCP} and \ref{fig:i45_LPCP}, respectively. 

The increases in the density and magnetic field strength lead to the polarization flips at 230~GHz. 
As shown in subsection \ref{subsec:jetLPreversal} and Appendix \ref{app:spectra}, the conditions for the flipping are satisfied at $\nu \sim \nu_{\rm c}$. 
Since $\nu_{\rm c} \propto B$, the typical flipping frequencies transit from $\sim$ 43 - 86~GHz to $\sim$ 130 - 250~GHz in the base regions of jets. 
In addition, the optical depth for SSA drastically increases due to the higher density and stronger magnetic fields. 

As a result, we see the synchronized LP and CP flips in both of low and high inclination cases. 
The images in Fig.~\ref{fig:i5_LPCP} show round-shaped, but non ring-like emissions due to the face-on-like observer. 
Both show the flipping on the counter jet side due to larger optical depths. 
In two panels in Fig.~\ref{fig:i45_LPCP}, the flipped components appear in the approaching jet, in addition to the counter jet. 
In this way, we can expect to detect the polarization flipping in the innermost region of diverse AGN jets through high resolution observations. 

\subsection{Comparison with Observations}\label{subsec:obs}

\begin{figure*}
\begin{center}
	\includegraphics[width=15cm]{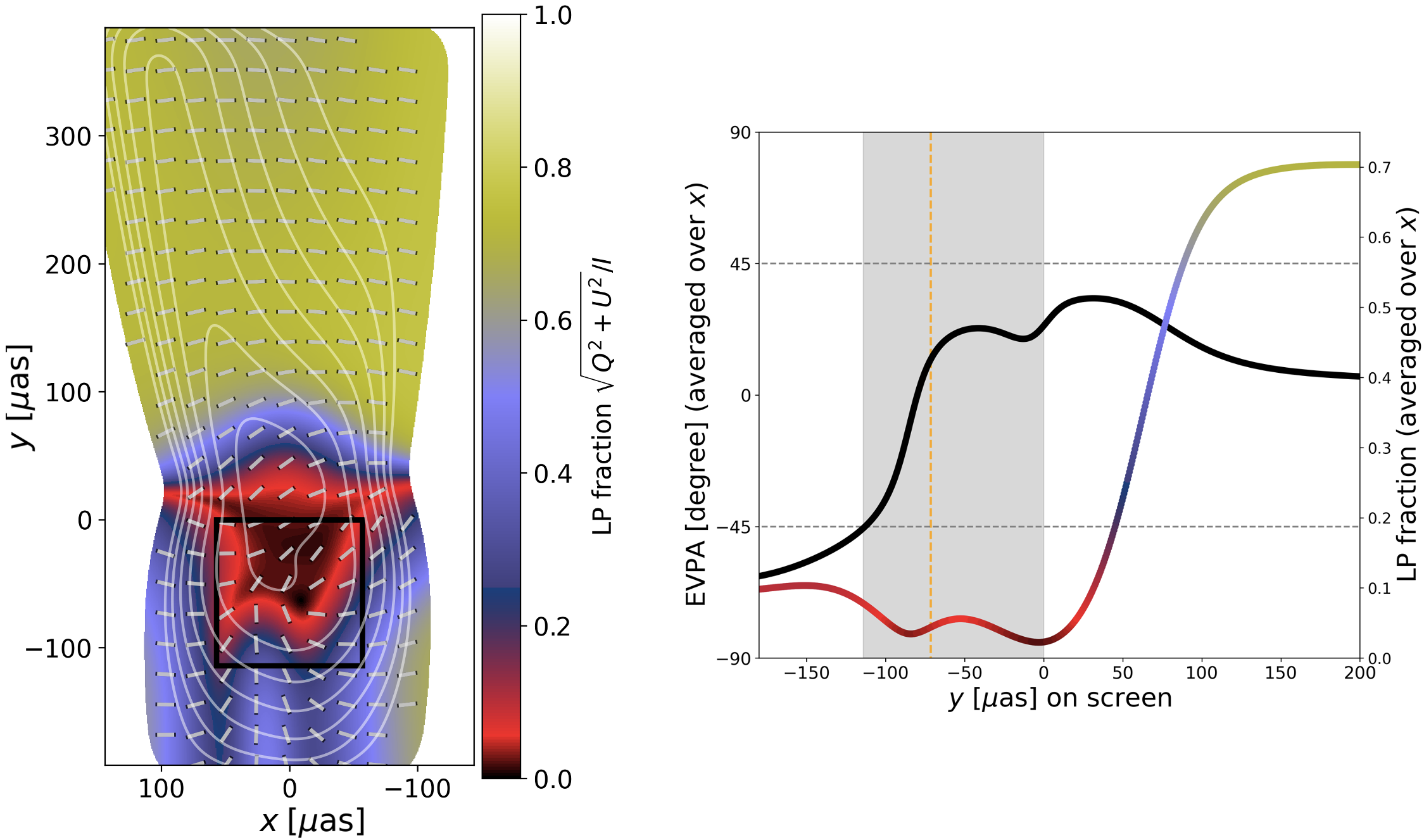}
\end{center}
    \caption{Left: Same as the left panel of Fig.~\ref{fig:43Gjet} but convolved image with $40~{\rm \mu as}$ circular Gaussian beam. (Here we have the M87 jet in mind.)  
    Right: Vertical profiles of the LP vector angle and LP fraction, obtained by horizontally averaging the left image. 
    The range of square in the left (gray) and the $\tau_{\rm SSA}=3$ point in the right of Fig.~\ref{fig:43Gjet} (orange dotted line) are overplotted. 
    }
    \label{fig:43Gjet_gauss}
\end{figure*}

Until here, we suggested that the flipping polarization components can be detected in the base regions of jets and on the photon ring. 
The sizes of the flipped regions are $\sim 10-20~r_{\rm g}$ in the jet base regions (e.g., Fig.~\ref{fig:43Gjet}) and $\sim 0.1~r_{\rm g}$ on the photon ring (Fig.~\ref{fig:86Gring}), respectively. 

If we assume the central black hole mass and distance of M87*, $r_{\rm g} \approx 3.8~{\rm \mu as}$ (see, for example, \cite{2019ApJ...875L...6E}), these regions extend on the scales of $\sim 40-80~{\rm \mu as}$ and $\sim 0.4~{\rm \mu as}$ on the sky. 
The former can be within the range of angular resolutions in near-future global VLBI observations, while the latter within the range of space-VLBI. For example, \citet{2023Natur.616..686L} accomplished the resolution of $\sim 40~{\rm \mu as}$ at 86~GHz and captured the innermost base region and ring-like feature of M87 jet. 

\bigskip

In Fig.~\ref{fig:43Gjet_gauss}, we convolve the LP map at 43~GHz (the left of Fig.~\ref{fig:43Gjet}) with circular Gaussian beam of $40~{\rm \mu as}$ to show as a mock-observational image, which is a little more ambitious than existing observations. 
The left panel illustrates a broadened jet feature and an LP vector pattern which exhibits smooth variation as a result of blurring. 

In order to examine the spatial variation of the position angle and fraction of LP vectors, we draw in the right panel the vertical profiles along $y$-axis after averaging over $x$-axis. 
Compared with the left panel of Fig.~\ref{fig:43Gjet} before convolving, it still shows an abrupt, but smoother and smaller change in EVPA in the flipping region. 
This might be, by itself, confused with the features caused by the Faraday rotation. 
However, we expect to detect the sign of LP flipping, if combined with the LP fraction. 

The profile of LP fraction in the right panel of Fig.~\ref{fig:43Gjet_gauss} shows two bounds at $y \sim -80~{\rm \mu as}$ and 0, which form a hump-like feature with values of $< 10\%$ at around $50{\rm \mu as}$. 
The first bound, going left to right, is synchronized with the nearly $90\degr$ flipping of EVPAs. 
The second bound corresponds to the transition from the counter to approaching jets. 
The hump feature is unique to the flipping by SSA, while the Faraday depolarization would give a smooth profile in the case that it is stronger in the inner region. 

The re-flipping feature as seen in the right of Fig.~\ref{fig:43Gjet} is absent here, because the re-flipping region is narrow and, hence, erased with the components from the approaching-side. 
(It should also be noted that the flipped vector pattern in the counter-side is similar to the unflipped pattern in the approaching-side.) 

\bigskip

Recently \citet{2020A&A...637L...6K} performed linearly polarimetric observations of M87 jet at 22 and 43~GHz, and reported the non-detection of the flip of LP vectors due to SSA (see their Appendix E). 
This is not surprising, since we have confirmed that the flipping LP vectors and CP components cannot be seen in the image convolved with a beam size of existing observations (e.g., $\sim 0.2~{\rm mas} = 200~{\rm \mu as}$). 
Meanwhile, they showed large, positive spectral indices in the base region of counter jet in their Fig.~B1, implying significant SSA effect in the counter side.
In this sense, our theoretical results are consistent with existing observations at the moment. 

Interestingly, their LP maps and those in \citet{2018ApJ...855..128W} show a flipping LP vector pattern between the approaching and counter jet sides in the base region (see also \cite{2021ApJ...922..180P}).
We expect that future observations with even higher resolution will be able to give a useful constraint on the existence (or absence) of $90\degr$-flip of LP vectors, thus shedding a new light on the particle population near the black holes.

\subsection{Other Triggers of Polarization Flips}
\label{subsec:ringLPvec}

Recently, several scenarios have been theoretically presented that give polarization flips on the photon ring. 
\citet{2022ApJ...929...49P} suggested that the direct $n=0$ emission and $n=1$ subring give opposite spiral LP vector patterns for poloidal magnetic field configuration. 
This gives a similar flipping feature to our results purely due to the light bending (see also \cite{2020PhRvD.101h4020H}). 

The LP flip in this work is different from theirs in that it is triggered by large opacity for SSA and is produced by radiations from the funnel jets. 
Thus, these two flipping mechanisms should be distinguishable through multiwavelength observations, since the SSA flipping is unique to low frequencies and disappears at high frequencies. 
In addition, the light-bending flip make a change in the handedness of spiral vector patterns, while the SSA flip give a $90\degr$ offset independent of the intrinsic vector pattern and magnetic field configuration.

For example, if both effects are present, the image at a low frequency gives similar \textit{spiral} vector patterns in the $n=0$ emission and $n=1$, twice-flipped, ring. However, observing at high enough frequency, one obtain a flipping \textit{spiral} vector pattern on the ring only by the light bending and poloidal magnetic fields. 

Meanwhile, if the $n=0$ emission gives a purely \textit{radial} (or \textit{azimuthal}) pattern, one obtain a SSA-flipped \textit{azimuthal} (\textit{radial}) pattern on the $n=1$ ring at a low frequency but no flipping at high frequencies (Daniel Palumbo, private communication). 

\bigskip

The CP flipping on the photon ring can also be given arise to by other effects than SSA. \citet{2021MNRAS.505..523R} suggested that the $n$-th sub-rings give flipping CP signs imprinting the twist of magnetic fields via Faraday conversion (see also \cite{2021MNRAS.508.4282M}). 

The flipping by SSA can be distinguished from this by the weaker CP fractions.
In our model, the SSA-flipping CP components at 86~GHz are presented with fraction of $\sim 0.3~\%$, while the flipping CPs in the optically thin case at 230~GHz give $\sim 1-2~\%$ in fraction.\footnote{We also confirmed in our model that the flipped CP components on the photon ring at 86~GHz show up even in the case turning off Faraday rotation and conversion and the angle effects of magnetic fields.} 

\bigskip

As for the flipping in jet, the LP vector pattern can be changed, depending on the relationship between magnetic field configuration and relativistic bulk motion of plasma. For example, the acceleration of jet can give a flip (vertical to horizontal, or vice versa) of vector pattern even with the magnetic field configuration being unchanged (see, for example, \cite{1981ApJ...248...87L,2013MNRAS.430.1504M}).

Nevertheless, as in the photon ring case, we can distinguish these change from the SSA-flipping by contrasting the same region in the optically thin and thick regimes at multi-wavelengths, and can verify the existence and distribution of nonthermal electrons in jet.

\section{Conclusions}

It has been pointed out in literatures that the LP vectors and CP components from optically thick, non-thermal synchortron plasmas show the flipping features; the $90\degr$-flipped LP vectors and the reversely rotating CP components, which are not seen in the optically thin case. 

To survey the feasibility of capturing these flipping polarizations and their relationship with the electron distribution, we performed GRRT calculations based on the semi-analytic, force-free jet models with realistic model parameters, bearing the M87 jet in mind. 

We find the following conditions for the non-thermal flipping components to be observed; 
\begin{itemize}
    \item The plasma around a black hole is optically thick for the SSA effect. Here, the criterion of thickness $\tau_{\rm SSA} \gtrsim 10$ is severer than the unpolarized case (Fig.~\ref{fig:IQrt}), while it can be loosened by Faraday effects (Appendix \ref{app:analysis}). 
    \item The SSA effect is dominant over the Faraday rotation and conversion effects. For the power-law distribution of electrons, it is typically satisfied at around the synchrotron characteristic frequency $\nu_{\rm c}$, at which the spectra of polarized SSA coefficients peak out (Fig.~\ref{fig:alphas_spectrum}).
    \item The flipping LP fraction is typically given by Eq.~\ref{eq:QoI}, but is decreased if with significant Faraday effects (Eq.~\ref{eq:QoIan}). In addition, at the frequency close to $\nu_{\rm c}$, it deviates from the fiducial value (Eq.~\ref{eq:thickLPf}) and decreases at low frequency (Fig.~\ref{fig:LP_spectrum}). 
    \item There should not exist bright, foreground optically thin sources between the optically thick plasma and the observer. Otherwise, since the flipping components are weak in the intensity, they can be drowned by the highly polarized optically thin emissions (Fig.~\ref{fig:map}). 
\end{itemize}

In the calculated images at millimeter wavelengths, the flipping LP and CP appear in the base region of counter jet and on the photon ring. 
The components from the counter-side are not intervened with the foreground emission. 
The light rays consisting of the photon ring earn large optical depths by the light bending. 
The flipping in the approaching-side jet can also be seen at low frequencies or in high inclination cases.

We furthermore confirmed that the flipping regions are synchronized with the high spectral index values $\alpha>2$, which are also an indicator of non-thermal SSA-thick plasma. 
In addition, the $90\degr$-flip of LP vectors gives rise to high RM values, which can lead to an additional degeneracy in observations. 
It was also verified that contribution from thermal disk does not break  the flipping features in our jet emission-dominant model. 
The polarization flipping over multi-wavelengths will give a constraint on physical quantities in the jet base, in combination with the analyses of spectral index and RM.

In the model with higher density and magnetic field strength, the polarization flipping can be seen in the images at 230~GHz. 
This is due to increase in optical depth and transition of $\nu_{\rm c} \propto B\gamma_{\rm min}^2$, and suggests the feasibility of survey of magnetic structure and particle composition in the innermost regions of various AGN jets. 

The predicted polarization flipping features can become accessible with near-future global-scale VLBI or more ambitious space-VLBI observations. 
The systematical comparison between the theoretical and observational results will provide us with a clue to understanding the creation and acceleration mechanisms of non-thermal particles in the jets.

\section*{Acknowledgements}

The authors thank Ramesh Narayan, Angelo Ricarte, and Daniel Palumbo for constructive comments and discussion and for providing the data for code comparison.
This work was supported in part by JSPS Grant-in-Aid for Early-Career Scientists JP18K13594 (TK), for Scientific Research (A) JP21H04488 (KO), and for Scientific Research (C) JP20K04026 (SM).
This work was also supported by MEXT as “Program
for Promoting Researches on the Supercomputer
Fugaku” (Structure and Evolution of the Universe Unraveled
by Fusion of Simulation and AI; Grant Number
JPMXP1020240219; TK, KO / Black hole accretion disks and quasi-periodic oscillations revealed by general relativistic hydrodynamics simulations and general relativistic radiation transfer calculations, JPMXP1020240054; TK, KO), by Joint Institute
for Computational Fundamental Science (JICFuS, KO),
and (in part) by the Multidisciplinary Cooperative Research Program in CCS, University of Tsukuba
(KO).
%KA is financially supported in part by grants from the National Science Foundation  (AST-1440254, AST-1614868, AST-2034306). 
Numerical computations were in part carried out on Cray XC50 at Center for Computational Astrophysics, National Astronomical Observatory of Japan. 

\onecolumn
\appendix
\renewcommand{\thesection}{\Alph{section}}

\section{Behaviors of Analytic Solutions and Contribution of Faraday Rotation}\label{app:analysis}

In Section \ref{subsec:eqs}, the typical scenario of flipping LP vector was outlined under a simplified case with synchrotron emission and SSA effect. 
Here, we consider a more general case taking Faraday rotation into account, 
\begin{equation}
	\frac{\rm d}{{\rm d}s}
	\left(
		\begin{array}{c}
			I \\
			Q \\
			U \\
		\end{array}
	\right)
		=
	\left(
		\begin{array}{c}
			j_I \\
			j_Q \\
			0 \\
		\end{array}
	\right)
		-
	\left(
		\begin{array}{ccc}
			\alpha_{I} & \alpha_{Q} & 0 \\
			\alpha_{Q} & \alpha_{I} & \rho_{V} \\
			0 & -\rho_{V} & \alpha_{I} \\
		\end{array}
	\right)
	\left(
		\begin{array}{c}
			I \\
			Q \\
			U \\
		\end{array}
	\right). 
	\label{eq:radtransIQU}
\end{equation}
While we again take $j_{U}=\alpha_{U}=0$, here $U$ can become non-zero due to the Faraday rotation $\rho_{V}$.

Eq.~\ref{eq:radtransIQU} can be analytically resolved in an analogous way with the unpolarized case ${\rm d}I/{\rm d}s = j_{I}-\alpha_{I}I$, through diagonalization of the $3 \times 3$ transfer matrix. 
The solutions are described as follows;
\begin{eqnarray}
	I\left(s\right) &=& \frac{\alpha_{Q}^{2}}{\alpha_{Q}^{2}-\rho_{V}^{2}} \left\{\frac{X_{+}+X_{-}}{2} -\left(\frac{\rho_{V}}{\alpha_{Q}}\right)^{2}X_{0}\right\} j_{I}% \nonumber \\
%	&& \ \ \ \ \ \	- \frac{\alpha_{Q}}{\sqrt{\alpha_{Q}^{2}-\rho_{V}^{2}}}\frac{X_{+}-X_{-}}{2} j_{Q}, \\
    - \frac{\alpha_{Q}}{\sqrt{\alpha_{Q}^{2}-\rho_{V}^{2}}}\frac{X_{+}-X_{-}}{2} j_{Q}, \\
	Q\left(s\right) &=& \frac{\alpha_{Q}}{\sqrt{\alpha_{Q}^{2}-\rho_{V}^{2}}}\frac{X_{+}-X_{-}}{2} j_{I} + \frac{X_{+}+X_{-}}{2} j_{Q}, \\
	U\left(s\right) &=& \frac{\alpha_{Q}\rho_{V}}{\alpha_{Q}^{2}-\rho_{V}^{2}}\left(X_{0}-\frac{X_{+}+X_{-}}{2}\right)j_{I}% \nonumber \\
%    && \ \ \ \ \ \ - \frac{\rho_{V}}{\sqrt{\alpha_{Q}^{2}-\rho_{V}^{2}}}\frac{X_{+}-X_{-}}{2} j_{Q}, 
    - \frac{\rho_{V}}{\sqrt{\alpha_{Q}^{2}-\rho_{V}^{2}}}\frac{X_{+}-X_{-}}{2} j_{Q}, 
\end{eqnarray}
where 
\begin{equation}
	X_{0} = \frac{1-{\rm exp}\left(-\alpha_{I}s\right)}{\alpha_{I}}, \
	X_{\pm} = \frac{1-{\rm exp}\left\{-\left(\alpha_{I}\pm\sqrt{\alpha_{Q}^{2}-\rho_{V}^{2}}\right)s\right\}}{\alpha_{I}\pm\sqrt{\alpha_{Q}^{2}-\rho_{V}^{2}}}. 
	\label{eq:Xs}
\end{equation}

The terms in Eq.~\ref{eq:Xs} express the decay and oscillation of solutions, and all converge into constant values, 
\begin{equation}
	X_{0} \rightarrow \frac{1}{\alpha_{I}}, \ % \\
	X_{\pm} \rightarrow \frac{1}{\alpha_{I}\pm\sqrt{\alpha_{Q}^{2}-\rho_{V}^{2}}}, 
\end{equation}
in the optically thick limit $\alpha_{I}s \gg 1$. 
More precisely, if $|\alpha_{Q}| > |\rho_{V}|$, the final convergence of solutions must wait that of the term $X_{-}$, which is satisfied for $(\alpha_{I}-\sqrt{\alpha_{Q}^{2}-\rho_{V}^{2}})s \gg 1$. 
In this sense, one can say that the radiation under polarized absorbers slowly converges for the same geological distance $s$, compared to that under the unpolarized ones. 
For example, under the SSA effect satisfying Eq.~\ref{eq:alphaLPf}, the convergence condition is $(\alpha_{I}-\alpha_{Q})s = \{4/(3p+10)\}\alpha_{I}s \gg 1$ if without Faraday rotation. 
Therefore, the solutions do not converge until $\alpha_{I}s \sim 30-40$, as seen in the plots in Fig.~\ref{fig:IQrt}.

\bigskip

In the optically thick limit, we obtain after calculation, 
\begin{equation}
	\left(\frac{Q}{I}\right)_{\rm thick} = \frac{\left\{1-\left(\frac{\rho_{V}}{\alpha_{Q}}\right)^{2}\right\}(j_{Q}\alpha_{I}-j_{I}\alpha_{Q})}
		{\left\{1-\left(\frac{\rho_{V}}{\alpha_{Q}}\right)^{2}\left\{1-\left(\frac{\alpha_{Q}}{\alpha_{I}}\right)^{2}\left(1-\left(\frac{\rho_{V}}{\alpha_{Q}}\right)^{2}\right)\right\}\right\}j_{I}\alpha_{I}-\left\{1-\left(\frac{\rho_{V}}{\alpha_{Q}}\right)^{2}\right\}j_{Q}\alpha_{Q}},
	\label{eq:QoIan}
\end{equation}
\begin{equation}
	\left(\frac{U}{Q}\right)_{\rm thick} = \frac{\rho_{V}}{\alpha_{I}}. 
	\label{eq:UoQan}
\end{equation}
If $|\rho_{V}| \ll |\alpha_{Q}|$, Eq.~\ref{eq:QoIan} can be written as 
\begin{equation}
	\left(\frac{Q}{I}\right)_{\rm thick} =  \frac{j_Q \alpha_I - j_I \alpha_Q}{j_I \alpha_I - j_Q \alpha_Q}\left\{1-\mathcal{O}\left(\left(\frac{\rho_{V}}{\alpha_{Q}}\right)^{2}\right)\right\}. 
	\label{eq:thickLPfapp}
\end{equation}
This coincides with Eq.~\ref{eq:QoI} for $\rho_{V}=0$. 

\bigskip

One also obtain an insight for another limit $|\rho_{V}| \gg |\alpha_{Q}|$.\footnote{Note that this corresponds to $|\rho_{V}s| \gg |\alpha_{Q}s| \gg 1$ here.} 
In this rotation-dominant case, Eqs.~\ref{eq:QoIan} and \ref{eq:UoQan} yields $(\sqrt{Q^{2}+U^{2}}/I)_{\rm thick} \ll 1$ and $(U/Q)_{\rm thick} \gg 1$. 
Thus, the LP vector converges into the position angle of $\pm 45\degr$ with respect to the projected magnetic field line, dependent on the sign of $\rho_{V}$, decaying in the fraction.

\section{Balance between SSA and Faraday effects}
\label{app:spectra}

\begin{figure*}
\begin{center}
	\includegraphics[width=12cm]{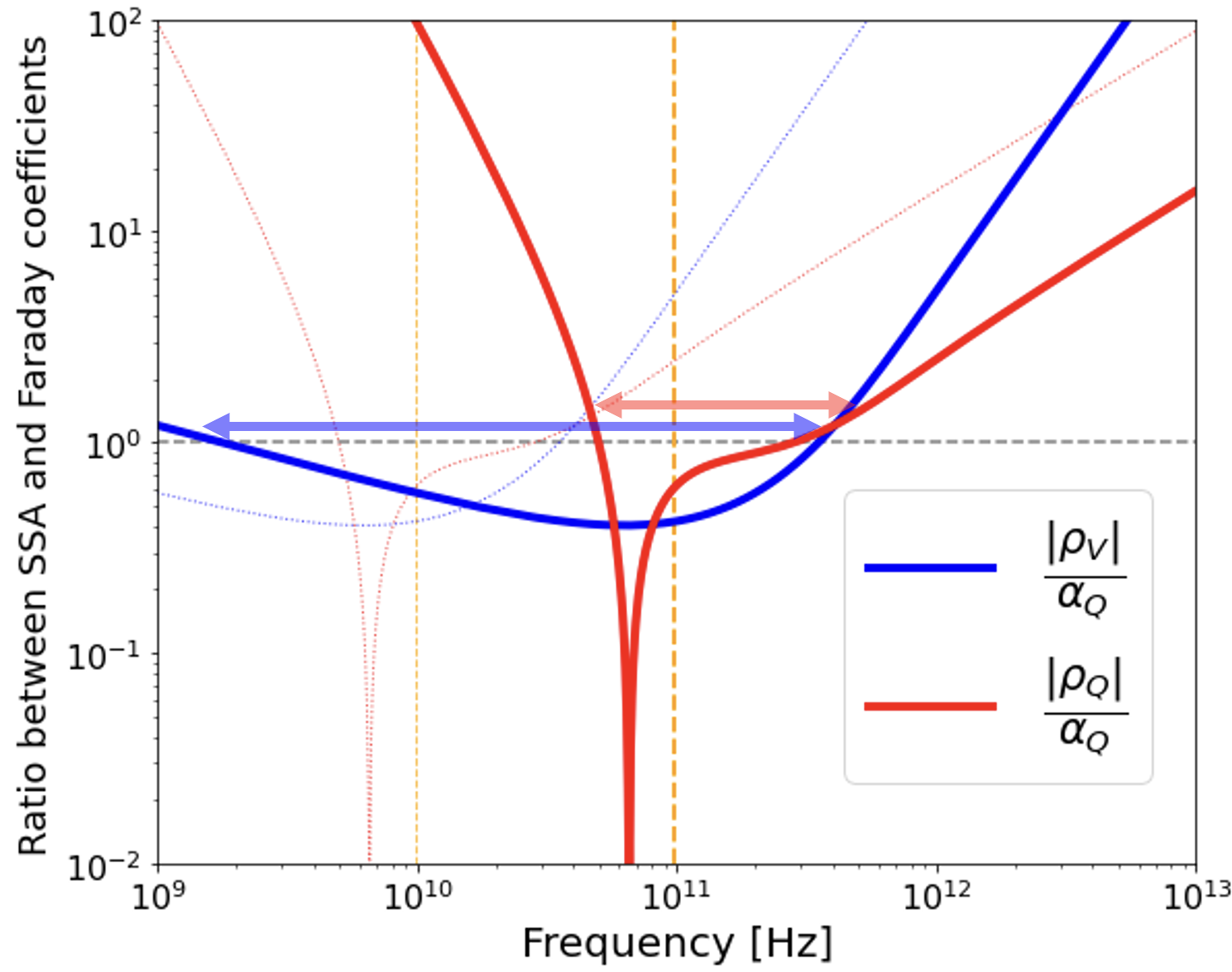}
\end{center}
    \caption{Diagram of the frequency-dependence of ratios between radiative coefficients for the SSA, $\alpha_{Q}$, and Faraday rotation $\rho_{V}$ and conversion $\rho_{Q}$ for the power-law electron distribution adopted in calculations in the main text, with $p=3.5$, $\gamma_{\rm min} = 30$ and $\gamma_{\rm max} = 10^{6}$. The thick solid and thin dotted lines correspond to the case with magnetic field strengths of $30~{\rm G}$ and $3~{\rm G}$, respectively. The angle between the directions of magnetic field and light path is assumed to be $\theta_{B} = 60\degr$. The orange dashed lines point the synchrotron characteristic frequency $\nu_{\rm c} = (3/2)\nu_{B}{\rm sin}~\theta_{B}\gamma_{\rm min}^{2}$ for the two magnetic field strengths. 
    Blue and red arrows point the frequency ranges on which the SSA dominates over Faraday rotation and conversion, respectively. The LP vectors and CP components can be flipped on these ranges.}
    \label{fig:alphas_spectrum}
\end{figure*}

In Appendix \ref{app:analysis}, we demonstrate that the sign-flipping of polarization components can be observed in the case where the SSA dominates over the Faraday effects. 
Here, we survey the applicability of this description under the physical condition in the AGN jets. 

In Fig.~\ref{fig:alphas_spectrum}, the coefficients for Faraday rotation and conversion, $\rho_{V}$ and $\rho_{Q}$, are plotted in fractions to the SSA coefficient $\alpha_{Q}$. 
The conditions of $|\rho_{V}| < |\alpha_{Q}|$ and $|\rho_{Q}| < |\alpha_{Q}|$, the blue and red lines below the gray dashed line, are simultaneously satisfied at a range around the synchrotron characteristic frequency $\nu_{\rm c} = (3/2)\nu_{B}{\rm sin}~\theta_{B}\gamma_{\rm min}^{2}$, at which the SSA coefficients peak out and dominate over the Faraday effects. 

The characteristic frequency $\nu_{\rm c}$ is located at millimeter to submillimeter wavelengths, of our interest, for the magnetic strength of $1 - 100~{\rm G}$ and the minimum Lorentz factor $\gamma_{\rm min} = 30$. 
%At the lower and higher frequencies than this range, the Faraday rotation and conversion effects are dominant over the SSA effect by orders of magnitude. 
Furthermore, it can be seen that the spectra show similar profiles for different magnetic field strength, scaled by $\nu_{\rm c} \propto B$. 
Since the values of magnetic field strength in our model are distributed in the innermost region of jets, $r \le 30 r_{\rm g}$, with a range of $B = 10 - 100~{\rm G}$, as  Fig.~\ref{fig:map}, the polarization flipping is obtained on the images at 43 and 86~GHz in the main text.

\bigskip

In the case that $|\rho_{V}| < |\alpha_{Q}|$ and $|\rho_{Q}| < |\alpha_{Q}|$ are satisfied, the LP fraction for the optically thick case $\alpha_{I}s \gg 1$ is approximately given by Eq.~\ref{eq:thickLPfapp}. 
It should also be noted here that the approximations of constant fractional polarizations, Eqs.~\ref{eq:jLPf}, \ref{eq:alphaLPf} and thus Eq.~\ref{eq:thickLPf}, are not necessarily applied at around $\nu_{\rm c}$, in which the spectra are affected by the minimum energy cutoff of power-law distribution at $\gamma_{\rm min}$. 

\begin{figure*}
\begin{center}
	\includegraphics[width=12cm]{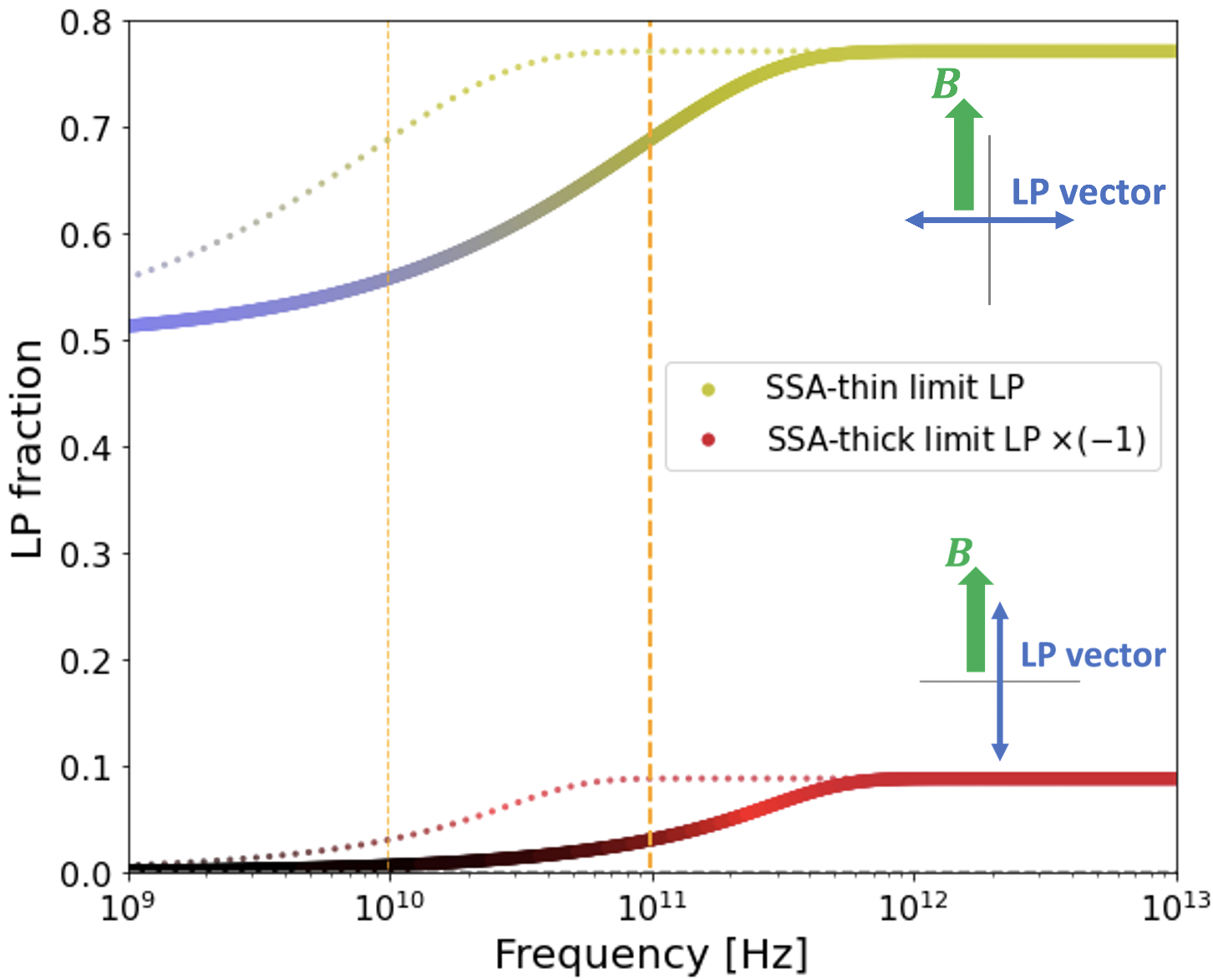}
\end{center}
    \caption{The frequency-dependence of LP fractions from an optically thin and thick, power-law synchrotron plasma. The solid and dotted lines describe the $30~{\rm G}$ and $3~{\rm G}$ cases, respectively, as in Fig.~\ref{fig:alphas_spectrum}. The color of lines corresponds to the color contour in Fig.~\ref{fig:43Gjet}.}
    \label{fig:LP_spectrum}
\end{figure*}

Fig.~\ref{fig:LP_spectrum} shows the LP fractions for the optically thick and thin cases at around $\nu_{\rm c}$. 
Both of the fractional values begin to deviate from and become lower than the constants of Eqs.~\ref{eq:jLPf} and \ref{eq:thickLPf}, which follows at the range of $\nu_{\rm c} \ll \nu \ll \nu_{\rm c}(\gamma_{\rm max}/\gamma_{\rm min})^{2}$. 
At the characteristic frequency $\nu_{\rm c}$, the SSA-thick LP fraction take a value of $\sim 5~\%$. %, which is consistent with the typical values obtained for the reversal cases in the main text. 
In the realistic calculations in the main text, the obtained LP fractions in the flipping cases are distributed in $\sim 0-10\%$ due to variable coefficients along the light paths.

\section{Images without SSA and Faraday Effects}
\label{app:noSSA}

\begin{figure*}
\begin{center}
	\includegraphics[width=12cm]{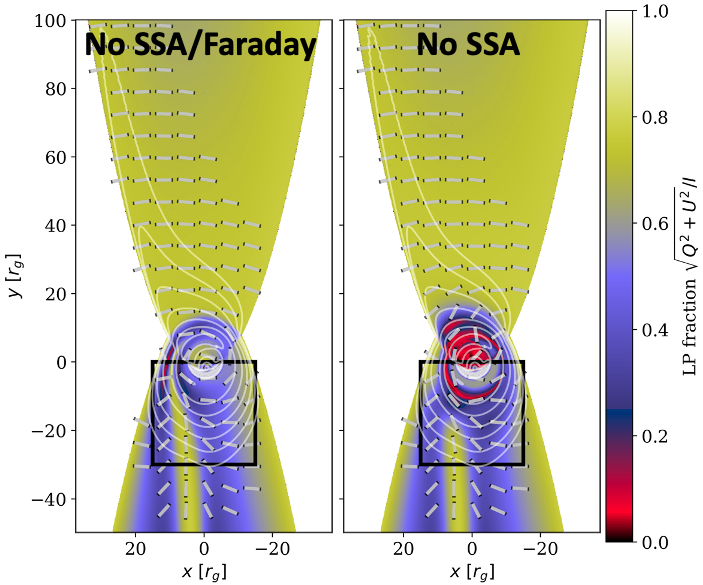}
\end{center}
    \caption{LP maps at 43~GHz in which we turn off both the SSA and Faraday effects (left) and in which turn off only SSA (right). The square in the left panels in Figs.~\ref{fig:43Gjet} is shown for comparison. 
    }
    \label{fig:noSSAnoF}
\end{figure*}

We show the LP maps at 43~GHz in which we turn off both of the SSA and Faraday rotation and conversion (left) and only SSA (right) in  Fig.~\ref{fig:noSSAnoF}. 
Even without the SSA, the LP vectors in the counter jet show a complicated pattern reflecting the magnetic field configuration. In the right panel, Faraday rotation gives rise to counter-clockwise offsets in EVPA in the base region relative to those in the left panel. 
However, the persistent $90\degr$-flips in LP vectors and decrease in LP flux $\lesssim 10\%$, as seen in Fig.~\ref{fig:43Gjet}, lacks due to the absence of SSA.

\section{GRRT Code Comparison}\label{app:codecomp}

In this work, we perform the polarization GRRT calculation with $\texttt{SHAKO}$ code, which we have developed and implemented in our previous works. 
To evaluate the consistence with other GRRT code, we here compare the polarization images obtained from two codes, $\texttt{SHAKO}$ and $\texttt{ipole}$. 
$\texttt{ipole}$ was firstly introduced in \citet{2018MNRAS.475...43M} and has been implemented in theoretical modelling of black hole images in the Event Horizon Telescope (EHT) Collaboration (e.g., \cite{2021ApJ...910L..13E}; see also \cite{2022ApJS..259...64W}). 

\begin{figure*}
\begin{center}
	\includegraphics[width=13cm]{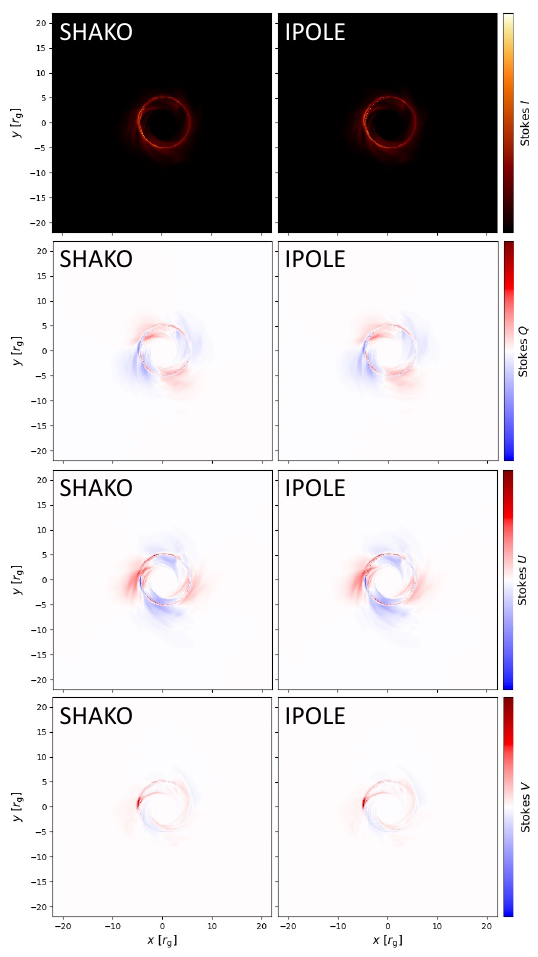}
\end{center}
    \caption{Image comparison for Stokes parameters $(I,Q,U,V)$ at 230~GHz obtained from GRRT calculations by two codes, $\texttt{SHAKO}$ (left column) and $\texttt{ipole}$ (right column). Both calculations are based on a MAD GRMHD model with black hole spin $a=0.9$ and electron temperature parameters $R_\mathrm{low} = R_\mathrm{high} = 1$. The observer's inclination angle is set to $i = 163\degr$.
    }
    \label{fig:shako_ipole}
\end{figure*}

In Fig.~\ref{fig:shako_ipole}, we show the images of Stokes parameters $(I,Q,U,V)$ at 230~GHz from $\texttt{SHAKO}$ (left column) and $\texttt{ipole}$ (right column). 
Both GRRT codes implement a general relativistic magnetohydrodynamics (GRMHD) simulation model for M87*, which belongs to magnetically arrested disk (MAD) regime (see \cite{2022MNRAS.511.3795N} for detail). 
Here, synchrotron emission, SSA, and Faraday rotation and conversion of thermal electrons are considered in GRRT calculation.
The black hole spin parameter is $a=0.9$ and two parameters in electron temperature prescription are set to $R_\mathrm{low} = R_\mathrm{high} = 1$ (R-$\beta$ prescription; \cite{2016A&A...586A..38M}). 
The observer screen's inclination angle is assumed to be $i = 163\degr$. 

All the images give a good agreement for two codes, showing the photon ring by gravity of the black hole and spiral-shaped radiations inside and outside of the ring from rotating plasma around the black hole. 
Since all the Stokes parameters are here affected significantly by all of SSA and Faraday rotation and conversion, it is demonstrated that the two codes are consistent in implementation of radiative coefficients and propagation of Stokes parameters in interaction with each other, in addition to in propagation of light paths and polarization reference planes around the black hole. 

In addition, we show the image-integrated values for two codes in Table~\ref{table:shako_ipole} for quantitative comparison. Here the total flux is $I_{\rm tot}$, the total LP and CP fractions are $\frac{\sqrt{Q_{\rm tot}^2+U_{\rm tot}^2}}{I_{\rm tot}}$ and $\frac{V_{\rm tot}}{I_{\rm tot}}$ respectively, the total EVPA is ${\rm arg}(Q_{\rm tot}+iU_{\rm tot})$, and the average LP fraction is $\frac{(\sqrt{Q^2+U^2})_{\rm tot}}{I_{\rm tot}}$. All the total values $(I_{\rm tot}, Q_{\rm tot}, U_{\rm tot}, V_{\rm tot})$, and $(\sqrt{Q^2+U^2})_{\rm tot}$ are calculated from the images in Fig.~\ref{fig:shako_ipole}. 
We again see a good agreement between two codes. All the deviation between two are within the code uncertainties in \citet{2023ApJ...950...35P}, in which the authors performed a comparison of polarization GRRT codes used in the EHT Collaboration.\footnote{the $\texttt{ipole}$ we use here corresponds to $\texttt{ipole-IL}$ in their labeling, which was introduced in \cite{2022ApJS..259...64W}.}

\bigskip

In conclusion, we confirmed the consistency of our code $\texttt{SHAKO}$ with $\texttt{ipole}$, and with the other several polarization GRRT codes adopted in calculation of images near black hole. 

\begin{table*}[]
\begin{center}
  \begin{tabular}{c|c|c|}
     & \texttt{SHAKO} & \texttt{ipole} \\ \hline \hline
    Total flux & 0.58375~Jy & 0.57953~Jy \\ \hline
    Total LP fraction & 2.5508~\% & 2.5643~\% \\ \hline
    Total CP fraction & 0.13183~\% & 0.13076~\% \\ \hline
    Total EVPA & $-4.5567\degr$ & $-4.2207\degr$ \\ \hline
    Average LP fraction & 45.035~\% & 45.206~\% \\ \hline
  \end{tabular}
\end{center}
  \caption{Comparison of image-integrated values from  Fig.~\ref{fig:shako_ipole} for two codes, $\texttt{SHAKO}$ (left column) and $\texttt{ipole}$ (right column). 
  \label{table:shako_ipole}
  }
\end{table*}

\newpage
\bibliographystyle{aasjournal}
\bibliography{nth_ref.bib}

\end{document}